\documentclass[10pt]{article}
\usepackage{float}
\usepackage[symbol]{footmisc} 
\usepackage[superscript,sort]{cite}
\usepackage{array}
\usepackage{bm}

\newcolumntype{x}[1]{%
>{\centering\hspace{0pt}}p{#1}}%
\usepackage[normalem]{ulem} 
\usepackage[colorlinks=true,linkcolor=black,citecolor=black,urlcolor=black,bookmarks=true,breaklinks=true]{hyperref}
\usepackage[usenames]{color}
\usepackage{amsmath,amssymb}

\usepackage{amsthm,amscd,amsxtra,amsfonts,amsmath,amssymb,multirow}
\usepackage{wrapfig}
\usepackage[footnotesize]{caption}
\usepackage{subcaption}
\usepackage[tiny,compact]{titlesec}
\usepackage{threeparttable} 
\usepackage{helvet}

\usepackage{booktabs}
\usepackage[table]{xcolor}
\usepackage{xcolor,colortbl}
\usepackage{placeins}
\usepackage{tikz}
\usetikzlibrary{shapes,arrows}
\usepackage[T1]{fontenc}
\usepackage[linesnumbered,ruled]{algorithm2e} 
\titlespacing*{\section}{0pt}{*0}{*0}
\titlespacing*{\subsection}{0pt}{*0}{*0}
\titlespacing*{\subsubsection}{0pt}{*0}{*0} 
\titlespacing{\paragraph}{0pt}{*0}{*1}

\definecolor{MyPurple}{rgb}{1,0,1}

\newcommand{\beq}[1]{\begin{equation} \label{#1}}
\newcommand{\eeq}{\end{equation}}
\newcommand{\barray}{\begin{array}{ll}}
\newcommand{\earray}{\end{array}}

\usepackage{longtable} 
\usepackage{threeparttable} 

\begin{document}
\pagenumbering{roman}

\clearpage \pagebreak \setcounter{page}{1}
\renewcommand{\thepage}{{\arabic{page}}}

\title{Representability of algebraic topology for biomolecules in machine learning based scoring and virtual screening}

\author{
Zixuan Cang$^1$,
Lin Mu$^2$,
 and
Guo-Wei Wei$^{1,3,4}$ \footnote{ Address correspondences  to Guo-Wei Wei. E-mail:wei@math.msu.edu}\\
$^1$ Department of Mathematics \\
Michigan State University, MI 48824, USA\\
$^2$ Computer Science and Mathematics Division \\
Oak Ridge National Laboratory, Oak Ridge, USA \\
$^3$  Department of Biochemistry and Molecular Biology\\
Michigan State University, MI 48824, USA \\
$^4$ Department of Electrical and Computer Engineering \\
Michigan State University, MI 48824, USA \\
}

\date{\today}
\maketitle
\maketitle
 \begin{abstract}  
This work introduces a number of algebraic topology approaches, such as multicomponent persistent homology, multi-level persistent homology and electrostatic persistence for the  representation,  characterization, and description of small molecules and biomolecular complexes. Multicomponent persistent homology retains critical chemical and biological information during the topological simplification of biomolecular geometric complexity. Multi-level persistent homology enables a tailored topological description of inter- and/or intra-molecular interactions of interest. Electrostatic persistence incorporates partial charge information into topological invariants. These topological methods are paired with Wasserstein distance to characterize similarities between molecules and are further integrated with a variety of machine learning algorithms, including k-nearest neighbors, ensemble of trees, and deep convolutional neural networks, to manifest their descriptive and predictive powers for chemical and biological problems. Extensive numerical experiments involving more than 4,000 protein-ligand complexes from the PDBBind database and near 100,000 ligands and decoys in the DUD database are performed to test respectively  the scoring power and the virtual screening power of  the proposed topological approaches. It is demonstrated that the present approaches outperform the modern machine learning based methods in protein-ligand binding affinity predictions and ligand-decoy discrimination.

 \end{abstract}

{\setcounter{tocdepth}{5} \tableofcontents}

\section{Introduction}

Arguably, machine learning has become one of the most important developments in data science and artificial intelligence. With its ability to extract features of various levels hierarchically, deep convolutional neural networks (CNNs) have made breakthroughs in image processing, video, audio, and computer vision~\cite{krizhevsky2012imagenet,simonyan2014very},  whereas recurrent neural networks  have found success in analyzing sequential data, such as text and speech \cite{lecun2015deep,hinton2012deep,schmidhuber2015deep,ngiam2011multimodal}.  Deep learning algorithms are able to automatically extract  high-level features and discover intricate patterns in large data sets. In general, one of major advantages of machine learning algorithms is their ability to deal with large data sets and uncover complicated relationships.  

Recently, machine learning  has become an indispensable tool in biomoelcular data analysis and structural bioinformatics. Almost every computational problem in  molecular biophysics and biology, such as the predictions of solvation free energy, solubility, partition coefficient,  protein-ligand  binding affinities, mutation induced protein stability change, molecular multipolar electrostatics, virtual screening etc,  has machine learning  based approaches that are either parallel or complementary to their physics based counterparts. The success of deep learning has fueled the rapid growth in several areas of biological science \cite{schmidhuber2015deep,lecun2015deep,ngiam2011multimodal}, including    bioactivity of small-molecule drugs  \cite{hughes2015modeling, unterthiner2015toxicity,lusci2013deep,wallach2015atomnet} and genetics \cite{dahl2014multi, ramsundar2015massively}, where large data sets are available.

Despite the success of deep CNNs in dealing with small molecules, the direct application of deep CNNs to three-dimensional (3D) macromolecule structures is extremely expensive and prone to accuracy reduction due to insufficient grid resolution and inadequate chemical labeling \cite{wu2017moleculenet}. As a result, deep CNNs have not been as competitive in terms of accuracy and efficiency as other commonly used machine learning algorithms, such random forest and gradient boosted trees, for predictions based on 3D biomolecular data. A major obstacle in the development of deep learning nets for 3D biomolecular data is  the presence  of geometric  and biological complexities. 

Biomolecules can be characterized by   geometric features,  electrostatic features, high level features, and amino-acid sequence features based on physical, chemical and biological understandings \cite{ZXCang:2017a}.  Geometric features, such as  coordinates, distances, angles, surface areas \cite{Bates:2008,Bates:2009,QZheng:2012}  and  curvatures \cite{ZhanChen:2010a,ZhanChen:2010b,ZhanChen:2012, DDNguyen:2016c}, are  important descriptors of biomolecules  \cite{XFeng:2012a, XFeng:2013b,KLXia:2014a}. However, geometric features  often involve too much structural detail and  are frequently computationally intractable for large biomolecular data sets.   Electrostatic features include atomic partial charges, Coulomb potentials, atomic electrostatic solvation energies, and polarizable multipolar electrostatics \cite{kandathil2013accuracy}. These descriptors become essential for highly charged biomolecular systems, such as nucleic acid polymers and some protein-ligand complexes. High level features refer to pKa values of ionizable groups and neighborhood amino acid compositions, such as the involvement of hydrophobic, polar, positively charged, negatively charged, and special case residues. Sequence features consist of secondary structures, position-specific scoring matrix (PSSM), and co-evolution information.  Sequence features and annotations provide a rich resource for bioinformatics analysis of biomolecular systems. 
Topology offers a new unconventional  representation of biomolecules. 
Topological features for biomolecules can be generated in a variety of ways \cite{KLXia:2014c}. Some of the most powerful topological features are obtained from multicomponent persistent homology, or element specific persistent homology (ESPH) \cite{ZXCang:2017a, ZXCang:2017b}.  Recently, we carried out a comprehensive comparison of  the performance of geometric features,  electrostatic features, high level features,  sequence features and topological features, for the prediction of  mutation induced protein folding free energy changes of four  mutation data sets \cite{ZXCang:2017a}. Surprisingly,  topological features  outperform all the other features \cite{ZXCang:2017a}.

Unlike geometry, topology is well known for its power of  simplification to  geometric complexity \cite{Schlick:1992trefoil,Zomorodian:2005,sumners:1992,IKDarcy:2013,CHeitsch:2014,Demerdash:2009,DasGupta2016,XShi:2011}. 
The global description generated by classical topology is based on the concept of neighborhood. If a space can be continuously deformed to another, they are considered to present the same topological features. In this sense, topology can not distinguish between a folded protein and its unfolded form. Such property prevents the use of classical topology  for the characterization of biomolecular structures. Instead of using topology to describe a single configuration of connectivity, persistent homology scans over a sequence of  configurations ordered by a filtration parameter and output a sequence of topological invariants, which partially captures part of geometric features. Persistent homology has been applied to biomolecular systems  in our earlier works.


In mathematics, persistent homology is a relatively new branch of algebraic topology \cite{Edelsbrunner:2002,Zomorodian:2005}. When dealing with proteins and small molecules, it is conventionally to consider  atoms as point clouds. For a given point cloud data set, one type of persistent homology turns each point into a sphere with their radii systematically increasing.  The corresponding topological invariants and their persistence over the varying  radius values can be computed. Therefore, this method embeds multiscale geometric information into topological invariants to achieve an  interplay between geometry and topology. Consequently,  persistent homology captures topological structures continuously over a range of spatial  scales. It is called persistent homology because at each given radius, topological invariants, i.e., Betti numbers,  are practically calculated by means of homology groups.  In the past decade, much theoretical formulation \cite{BH11,CEH07,CEH09,CEHM09,CCG09,CGOS11,Carlsson:2009theory,CSM09,SMV11,zigzag} and  many computational algorithms \cite{OS13,DFW14,Mischaikow:2013,javaPlex,Perseus, Dipha} have been  developed. One-dimensional (1D) topological invariants generated from persistent homology is often  visualized by persistence barcodes \cite{CZOG05,Ghrist:2008} and  persistence diagrams \cite{edelsbrunner:2010}. In recent years, multidimensional persistence has attracted much attention  \cite{Carlsson:2009computing,Carlsson:2009theory} in hope that it can better characterize the data shape when there are multiple measurements of interest.

Persistent homology  has been applied to  various fields, including image/signal analysis \cite{Carlsson:2008,Pachauri:2011,Singh:2008,Bendich:2010,Frosini:2013,Perea:2015a}, chaotic dynamics verification \cite{Mischaikow:1999,kaczynski:mischaikow:mrozek:04}, sensor networks \cite{Silva:2005}, complex networks \cite{LeeH:2012,Horak:2009}, data analysis \cite{Carlsson:2009,Niyogi:2011,BeiWang:2011,Rieck:2012,XuLiu:2012}, shape recognition \cite{DiFabio:2011,AEHW06,Feng:2013} and computational biology \cite{Kasson:2007,Gameiro:2014,Dabaghian:2012,Perea:2015b,Gameiro:2014}.  Compared with traditional computational topology \cite{Krishnamoorthy:2007,YaoY:2009,ChangHW:2013}  and/or computational homology, persistent homology  inherently adds an additional dimension, i.e., the filtration parameter. The filtration parameter can be used to embed important geometric or quantitative information into topological invariants. As such, the importance of retaining  geometric information in topological analysis has been recognized  \cite{Biasotti:2008}, and  persistent homology  has been advocated as a new approach for handling big data sets \cite{BVP15, BHPP14,Fujishiro:2000,Carlsson:2009,Ghrist:2008}. Recently, we have introduced persistent homology for mathematical modeling and prediction of  nano-particles, unfolding proteins and other biomolecules \cite{KLXia:2014c, KLXia:2015a}.  We proposed the molecular topological fingerprint (TF)   to reveal topology-function relationships in protein folding and protein flexibility \cite{KLXia:2014c}. 
We established some of the first quantitative topological analyses in our persistent homology based  predictions of  the curvature energy of fullerene isomers  \cite{KLXia:2015a,BaoWang:2016a}. We have also shown correlation between persistence barcodes and energies computed with physical models during molecular dynamics experiments \cite{KLXia:2014c}.  Moreover, we have introduced the first differential geometry based persistent homology that utilizes partial differential equations (PDEs) in filtration   \cite{BaoWang:2016a}. 
Most recently, we have developed a topological representation to address additional measurements of interest, by stacking the persistent homology outputs from a sequence of frames in molecular dynamics or a sequence of different resolutions \cite{KLXia:2015d,KLXia:2015e}.
We have also introduced the first use of topological fingerprints  for resolving ill-posed inverse problems in  cryo-EM  structure determination \cite{KLXia:2015b}. In 2015, we  constructed one of  the first topology based machine-learning algorithms in protein classification involving tens of thousands of proteins and hundreds of tasks \cite{ZXCang:2015}.  We also developed   persistent-homology based software for the automatic detection of protein cavities and binding pockets \cite{ESES:2017}. 

Despite of much success, it was found that persistent homology has a limited characterization power for proteins and protein complexes, when applied directly to a single selection of atoms \cite{ZXCang:2015}. Essentially, biomolecules are not only complex in their geometric constitution, but also intricate in biological constitution. In fact, the biological constitution is essential to biomolecular structure and function. Persistent homology that is designed to reduce the geometric complexity of a biomolecule can easily miss biological information. To overcome this difficulty, we have introduced multicomponent persistent homology or element specific persistent homology (ESPH) to recognize the chemical constitution during the topological simplification of biomolecular geometric complexity  \cite{ZXCang:2017a, ZXCang:2017b,ZXCang:2017c}.  In ESPH,  the atoms of a specific set of element types in a biomolecule are selected so that certain chemical information is emphasized.    Our  ESPH is not only able to outperform other geometric and electrostatic representations in large data sets, but is also able to shed light on the molecular mechanism of protein-ligand binding, such as the relative importance of  hydrogen bond,  hydrophilicity and hydrophobicity at various spatial ranges \cite{ZXCang:2017b}.    

The objective of the present work is to further explore the representability
 and reduction power of multicomponent persistent homology  for biomolecules.  To this end, we take a combinatorial approach to scan a variety of element combinations and examine the characterization power of these components. Additionally, we also propose a multi-level 
persistence to study the topological properties of non-covalent bond interactions. This approach enables us to devise persistent homology to describe the interactions of interest between atoms that are connected by a series of covalent bonds, and delivers richer representation especially for small molecules.   
Moreover, to enhance the power of topological representation, we introduce electrostatic persistence, which embeds charge information in topological invariants, as a new class of features in multicomponent persistent homology.  Electrostatics are of paramount importance in biomolecules. 
The aforementioned approaches can be realized via the modification of the distance matrix with a more abstract setting, for example, Vietoris-Rips complex. 
The complexity reduction is guaranteed in the 1D  topological  representation of 3D biomolecular structures.   Obviously, the multicomponent persistent homology representation of biomolecule leads to a higher  machine learning dimensionality compared to the original single component persistent homology for a biomolecule. Therefore, it  is subject to overfitting or overlearning problem in machine learning theory. 
Fortunately, gradient boosting trees (GBT) method is relatively insensitive to redundant high dimensional topological features \cite{ZXCang:2017a}.  Finally, since components can be arranged  as a new dimension ordered by their feature importance,  multicomponent  persistent homology barcodes are naturally a two-dimensional (2D) representation of biomolecules. Such a 2D representation can be easily used as image-like input data in a deep CNN architecture, with different topological dimensions, i.e., Betti-0, Betti-1, and Betti-2, being treated as multichannels. Such approach addresses the nonlinear interactions among important element combinations while keeping the information from less important ones. Barcode space metrics, such as bottleneck distance  and more generally, Wasserstein metrics \cite{cohen2010lipschitz,carlsson2014topological}, offer a direct description of similarity between molecules and can be readily used with nearest neighbor regression or kernel based methods.  The performance of  Wasserstein distance for protein-ligand binding affinity predictions is examined in this work.
 
The rest of this manuscript is organized as follows. Section \ref{Methods} is devoted to introduce methods and algorithms.  We present multicomponent persistent homology, multi-level interactive persistent homology, vectorized persistent homology representation and electrostatic persistence. These formulations are crucial for the representability of persistent homology for biomolecules. Machine learning algorithms associated  with the present topological data analysis are briefly discussed. Results are presented in Section \ref{Results}. We first consider the characterization of small molecules. More precisely, the cross-validation of protein-ligand binding affinities prediction via solely ligand topological fingerprints is studied. We illustrate the excellent representability of our multicomponent persistent homology by a comparison with physics based descriptors. 
Additionally, we investigate representational power of the proposed topological method  on  a few sets of benchmark  protein-ligand binding data sets, namely, PDBBind v2007, PDBBind v2013, PDBBind v2015 and BPDBind v2016 \cite{PDBBind:2015}. These data sets contain thousands of protein-ligand complexes and have been extensively studied in the literature. Results indicate that multicomponent persistent homology offers one of most powerful representations of protein-ligand binding systems. 
The aforementioned study of the characterization of small molecules and protein-ligand complexes leads to an optimal selection of features and models to be used for virtual screening.
Finally, we consider the directory of useful decoys (DUD) database  to examine the representability of our multicomponent persistent homology for virtual screening to distinguish actives from decoys. The DUD data set used in this work  has a total of about 100000 compounds containing 2950 active ligands, which  bind  to 40 targets from six families. A large number of  state-of-the-art virtual screening methods have been applied to this data set. We demonstrate that the present  multicomponent persistent homology outperforms all other methods. This paper ends with a conclusion.

\section{Methods and algorithms}\label{Methods}

\subsection{Persistent homology} 
The concept of persistent homology is built on the mathematical concept of homology, which associates a sequence of algebraic objects, such as abelian groups, to topological spaces. For discrete data such as atomic coordinates in  biomolecules,
algebraic groups can be defined via simplicial complexes, which are constructed from simplices, generalizations of the geometric notion of nodes, edges, triangles and tetrahedrons to arbitrarily high dimensions. Homology characterizes the topological connectivity of geometric objects in terms of topological invariants, i.e., Betti numbers, which are used to distinguish topological spaces by counting $k$-dimensional holes.  Betti-0, Betti-1 and Betti-2, respectively, represent independent components, rings and cavities in a physical sense. In persistent homology, Betti numbers are evaluated along with a filtration parameter, such as the radius of a ball or the level set of a hypersurface function,  that continuously varies over an interval. Therefore, persistent homology is induced by the filtration. For a given biomolecule, the change and the persistence of topological invariants over the filtration offer a unique characterization.  These concepts are very briefly discussed below.

\subsubsection{Simplicial complex}
 
\paragraph{Simplex} 
A \textit{k-simplex} denoted by $\sigma^k$ is the convex hull of $k+1$ affinely independent points in $\mathbb{R}^k$. The convex hull of each nonempty subset of the $k+1$ points forms a subsimplex and is regarded as a \textit{face} of $\sigma^k$. The points are also called \textit{vertices} of $\sigma^k$.

\paragraph{Simplicial complex} 

A set of simplices $K$ is a \textit{simplicial complex} if all faces of any simplex in $K$ are also in $K$ and the intersection of any pair of simplices in $K$ is either empty or a common face of the two simplices.

\subsubsection{Homology}\label{homology}
 
\paragraph{Chain} A \textit{k-chain} of a simplicial complex $K$ denoted by $C_k(K)$ is the formal linear combination of all the \textit{k-simplices} in $K$. Here, we take the $\mathbb{Z}_2$  field for the coefficients of the linear combination. Under the rule of fields, a \textit{k-chain} is a group called \textit{chain group}.

\paragraph{Boundary operator} A \textit{boundary operator} denoted by $\partial_k:C_k(K)\rightarrow C_{k-1}(K)$ maps a linear combination of \textit{k-simplices} to the same linear combination of the boundaries of the \textit{k-simplices}. With a \textit{k-simplex} $\sigma^k=[v_0,\ldots, v_k]$ where $v_i$ are the vertices of $\sigma^k$, the \textit{boundary operator} is defined as $\partial_k\sigma^k = \sum\limits_{i=0}^{k}\sigma^{k-1}_i$, where $\sigma^{k-1}_i$ is a \textit{(k-1)-simplex} which is a face of $\sigma^k$ with the $i$th vertex being absent.

\paragraph{Cycle group} A \textit{k-cycle} is a \textit{k-chain} whose image under the \textit{boundary operator} $\partial_k$ is the empty set. The collection of all the \textit{k-cycles} forms a group denoted by $Z_k(K)$ which is the kernel of $\partial_k:C_k(K)\rightarrow C_{k-1}(K)$.

\paragraph{Boundary group} The image of $\partial_k+1: C_{k+1}(K)\rightarrow C_k(K)$ is called the boundary group and is denoted by $B_k$. $B_k$ is a subgroup of $Z_k$ following the property of the \textit{boundary operator} that $\partial_{k}\circ\partial_{k+1} = 0$.

\paragraph{Homology group} The $k$th \textit{homology group} is the quotient group defined as $H_k=Z_k/B_k$. It is used to compute Betti numbers.

\paragraph{Betti number} The $k$th \textit{Betti number} $\beta_k$ is defined and often computed as $\text{rank}H_k = \text{rank}Z_k-\text{rank}B_k$.

\subsubsection{Persistent homology}

\paragraph{Filtration} A \textit{filtration} of a \textit{simplicial complex} $K$ is a nested sequence of subcomplexes of $K$ such that $\varnothing=K^0\subset K^1 \subset \ldots\subset K^m=K$. Each $K^i$ is itself a \textit{simplicial complex}.

\paragraph{Persistence} Given a \textit{simplicial complex} $K$ with its filtration $K^i$, the \textit{p-persistent $k$th homology group} of $K^i$ is defined as $H_k^{i,p} = Z_k^i/(B_k^{i+p}\cap Z_k^i)$.

\subsubsection{Simplicial complexes and filtration}
Given a finite set of points $X$ and a non-negative scale parameter $r$, the Vietoris-Rips complex and alpha complex are constructed as follows.

\paragraph{Vietoris-Rips complex}
With a predefined metric $d(\cdot,\cdot)$ in $X$, a subset $X'$ of $X$ forms a simplex if $d(x_i,x_j)\leq r \, \text{for} \, \text{all} \, x_i,x_j \in X'$. The collection of all such simplices is the Vietoris-Rips complex of the finite metric space $X$ with scale parameter $r$ denoted by $Rips(X,r)$. It is obvious that $Rips(X,r)\subseteq Rips(X,r')$ for $r\leq r'$.

\paragraph{Alpha complex}
With $Alpha(X,r)$ being the alpha complex of $X$ with the scale parameter $r$ and given the Delaunay triangulation induced by the Voronoi diagram of $X$, a simplex in the Delaunay triangulation belongs to $Alpha(X,r)$ if all its \textit{1-faces} (\textit{1-simplex} as subset of the simplex) have length no greater than $2r$. Similar to Rips complex, alpha complex also has the property that $Alpha(X,r)\subseteq Alpha(X,r')$ for $r\leq r'$.

\subsection{Biological considerations}

The development of persistent homology was motivated by its potential in the dimensionality reduction, abstraction and simplification of biomolcular complexity \cite{Edelsbrunner:2002}. In the early applications of persistent homology to biomolecules, emphasis was given on  major or global features. Short barcodes were regarded noise.  For example, persistent homology  was used to identify the tunnel in a Gramicidin A channel \cite{Edelsbrunner:2002} and to study membrane fusion \cite{kasson2007persistent}. 
For the predictive modeling of biomolecules, features of a wide range of scales might all be important to the target quantity \cite{KLXia:2014c}. In the global scale, the biomolecular conformation should be captured. In the intermediate scale, the smaller intradomain cavities need to be identified. In the most local scale, the important substructures should be addressed, such as the pyrrolidine in the side chain of proline. These features of different scales can be reflected by barcodes with different centers and persistences. Therefore, applications in biomolecules can make a more exhaustive use of persistent homology \cite{KLXia:2014c,KLXia:2015a}, compared to some other applications where only global features matter while most local features are mapped to noise. Earlier use of persistent homology was focused on qualitative analysis. Only recently had persistent homology been devised as a quantitative tool \cite{KLXia:2014c,KLXia:2015a}. 
While the aforementioned applications are descriptive and regression based analysis, we have also applied persistent homology to predictive modeling of biomolecules \cite{ZXCang:2015}. However, biomolecules are both structurally and biologically complex. Their  geometric and biological complexities include covalent bonds, non-covalent interactions, effects of chirality, cis and trans distinctions,  multi-leveled protein structures,  and  protein-ligand and protein-nucleic acid complexes. Covering a large range of spatial scales is not enough for a power model. The biological details should also be explored. We address the underlying biology and physics by modifying the distance function and selecting various sets of atoms according to element types, to describe different interactions. Some biological considerations are discussed in this section.


\paragraph{Covalent bonds}

Covalent bonds are formed via  shared electron pairs or bonding pairs. The lengths and the number of covalent bonds can be easily detected from Betti-0 barcodes. For macromolecules, the same type of covalent bonds have very similar bond lengths and thus Betti-0 barcode patterns.  

\paragraph{Non-covalent interactions}
Non-covalent interactions play a critical role in maintaining the 3D structure of biomolecules and mediating chemical and biological processes, such as solvation, partition coefficient,  binding,  protein-DNA specification, molecular self-assembly, etc.  Physically, non-covalent interactions are due to  electrostatic,  van der Waals forces, hydrogen bonds, $\pi$-effects, hydrophobic effects, etc. The ability to characterize non-covalent interactions is an essential task in any methodological development. Betti-1 and Betti-2 barcodes are suitable for the characterization of the arrangement of such interactions in a larger scale. Additionally, we propose multi-level persistence and electrostatic persistence to reveal local and pairwise non-covalent interactions via Betti-0 barcodes as well. 


\paragraph{Multi-leveled protein structures}
Protein structures are typically described in terms of primary, secondary,  tertiary and quaternary ones. The protein primary structure is the linear sequence of amino acids in the polypeptide chain. A secondary structure mainly refers to $\alpha$-helix and $\beta$-sheets, which are highly regular and can be easily detected by distinct  Frenet-Serret frames.  A tertiary structure refers to the 3D structure of a single polypeptide chain. Its formation involves various non-covalent interactions, salt bridges, hydration effects, and often disulfide bonds.   A quaternary structure refers to  the aggregation of two or more individual polypeptide chains into a 3D multiprotein complex. Protein structures are further complicated by its functional domains, motifs, and particular folds. The protein structural variability and  complexity result in the challenge and opportunity for  methodological developments.  Various persistent homology techniques, including  multicomponent, multi-level, multidimensional \cite{KLXia:2015c}, multiresolution  \cite{KLXia:2015e},  electrostatic,  and interactive \cite{ZXCang:2017b} persistent homologies have been designed either in our earlier work or in this paper for protein  structural variability and  complexity. 

\paragraph{Protein-ligand, protein-protein, and protein-nucleic acid complexes}
Topological characterization of proteins is further complicated by protein interactions or binding with ligands (drugs), proteins, DNA and/or RNA molecules. Although a normal protein involves only carbon (C), hydrogen (H), nitrogen (N), oxygen (O) and sulfur (S) atoms, its protein-ligand complexes bring a variety of other elements into the play, including,   phosphorus (P), fluorine (F), chlorine (Cl), Bromine (Br), iodine (I), and many important biometals, such as calcium (Ca), potassium (K) sodium (Na), iron (Fe), copper (Cu), cobalt (Co), zinc (Zn), manganese (Mn), chromium (Cr), vanadium (V), tin (Sn), and molybdenum (Mo). Each biological element has important biological functions and its presence in biomolecules should be treated uniformly as a set of  points in the point cloud data. The interaction of protein and nucleic acids can be very intricate. Qualitatively, multiscale and multi-resolution persistent homology demonstrates interesting features in 3D DNA structures \cite{KLXia:2015d}. Typically, 3D RNA structures are more flexible and difficult to extract topological patterns. Interactive persistent homology, element specific persistent homology and binned representation for persistent homology outputs were designed to deal with interactions between protein-ligand, protein-protein,  and protein-nucleic acid  complexes \cite{ZXCang:2017a,  ZXCang:2017b, ZXCang:2017c }.  These approaches worked well in protein-mutation site interactions \cite{ZXCang:2017a}.   Additionally, multi-level persistent homology and electrostatic persistence proposed in this work are useful tools to describe some other specific interactions.

\subsection{Element specific persistent homology}
One important issue is how to  protect chemical and biological information during the topological simplification. As mentioned early, one should not treat different types of atoms as structureless points in a point cloud data. To this end, element specific persistent homology or multi-component persistent homology has been proposed to retain biological information in topological analysis \cite{ZXCang:2017a,ZXCang:2017b,ZXCang:2017c}. The element selection is similar to a predefined vertex color configuration for graphs. 

When all atoms are passed to persistent homology algorithms, the information extracted mainly reflects the overall geometric arrangement of a biomoelcule at different spatial scales. By passing only atoms of certain element types or of certain roles to the persistent homology analysis, different types of interactions or geometric arrangements can be revealed. In  protein-ligand binding modeling, the selection of all carbon atoms characterizes the hydrophobic interaction network whilst the selection of all nitrogen and/or oxygen atoms characterizes hydrophilic network and the network of potential hydrogen bonds. In the protein structural analysis, computation on all atoms can identify geometric voids inside the protein which may suggest structural instability and computation on only C$_\alpha$ atoms reveals the overall structure of amino acid backbones.   In addition, combination of various selections of atoms  based on element types provides very detailed description of the biomolecular system and hidden relationships from the structure to function can then be learned by machine learning algorithms. This may lead to the discovery of important interactions not realized as \textit{a prior}. This can be realized by passing the set of atoms of the selected element types to the persistent homology computation. This concept is used with the various definitions of distance matrix discussed as follows.


\subsection{Distance matrix induced persistent homology} 
Biomolecular systems are not only complex in geometry, but also in chemistry and biology. To effective  describe complex biomolecular systems, it is necessary to modify the filtration process. There are three commonly used filtrations, namely, radius filtration, distance matrix filtration, and density filtration, for biomolecules \cite{KLXia:2014c, KLXia:2015e}.  
A distance matrix defined with smoothed cutoff functions was proposed in our earlier work to deal with interactions within a spatial scale of interest in biomolecules \cite{KLXia:2014c}. In the present work, we introduce more distance matrices to enhance the representational power of persistent homology and to cover some important interactions that were not covered in our earlier works. The distance matrices can be used with a more abstract construction of simplicial complexes, such as Vietoris-Rips complex.

\subsubsection{Multi-level persistent homology}

Small molecules such as ligands in protein-ligand complexes usually contain fewer atoms than large biomolecules such as proteins. Bonded atoms stay closer than non-bonded ones in most cases. As a result,  the collection of Betti-0 bars will mostly provide the information about the length of covalent bonds. It is difficult to capture non covalent bond interactions among atoms especially  hydrogen bonds and  van der Waals pairwise interactions  in  Betti-0 barcodes. In order to describe non covalent interactions , we propose multi-level persistent homology, by simply modifying the distance matrix, similar to the idea of modifying distance matrix to emphasize on the interactions between protein and ligand \cite{ZXCang:2017b}. Given the original distance matrix $\mathbf{M}=(d_{ij})$ with $1\leq i,j\leq N$, the modified distance  matrix is defined as
\begin{equation}\label{eq:order1matrix}
\mathbf{\widetilde{M}}_{ij} = 
\begin{cases}
d_\infty,\,\rm{if\, atoms}\, $i$ \,\rm{and}\, $j$ \,\rm{are \,bonded}, \\
\mathbf{M}_{ij},\, \rm{otherwise},
\end{cases}
\end{equation}
where $d_\infty$ is a large number which is set to be greater than the maximal filtration value chosen by a persistent homology algorithm. Note that this matrix may fail to satisfy triangle inequality whilst still satisfies the construction principle of Rips complex.  

\begin{figure}[ht]
\begin{center}
\includegraphics[scale=0.4]{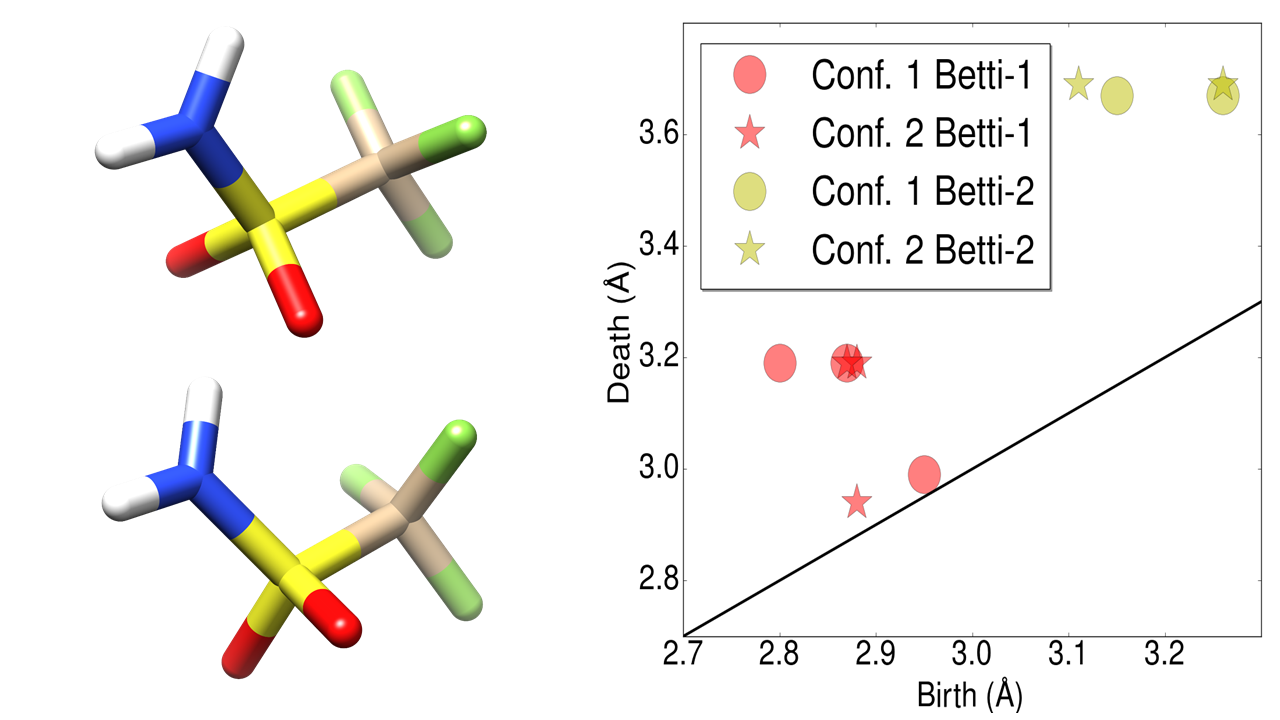}
\end{center}
\caption{Illustration of representation ability of $\mathbf{\widetilde{M}}$ in reflecting structural perturbations among conformations of the same molecule. \textbf{a}. The structural alignment of  two conformations of the ligand in protein-ligand complex (PDB:1BCD). \textbf{b}. The persistence diagram showing the Betti-1 and Betti-2 results generated using Rips complex with $\mathbf{\widetilde{M}}$ for  two conformations. It is worth noticing that the barcodes generated using Rips complex with $\mathbf{M}$ are identical for the two conformations.}\label{fig:order1illustration}
\end{figure}

The present multi-level persistent homology is able to describe any selected interactions of interest and delivers two benefits in characterizing biomolecules. Firstly, the pairwise non-covalent interactions can be reflected by Betti-0 barcodes. Secondly, such treatment generates more higher dimensional barcodes and the small structural fluctuation among different conformations of the same molecule can be captured. The persistent barcode representation of the molecule can be significantly enriched to better distinguish between different molecular structures and isomers. As an illustration, we take the ligand from the protein-ligand complex with PDB code ``1BCD" which only has 10 atoms. A different conformation of the ligand is generated by using the Frog2 server \cite{miteva2010frog2}. The persistent barcodes generated using Rips complex with the distance matrices $\mathbf{M}$ are identical and only have Betti-0 bars due to the simple structure. In this case,  Betti-0 bars only reflect the length of each bond and therefore fail to distinguish two slightly different conformations of the same molecule. However, when the modified distance matrices $\mathbf{\widetilde{M}}$ are employed, the barcode representation is significantly enriched and is able to capture the tiny structural perturbation between conformations. An illustration of the outcome from the modified distance matrix $\mathbf{\widetilde{M}}$ is shown in Figure \ref{fig:order1illustration}. A general $n$th level persistence characterization of molecules can be obtained with the distance matrix $\mathbf{\widetilde{M}}^n$ as,
\begin{equation}\label{eq:ordernmatrix}
\mathbf{\widetilde{M}}^n_{ij} = 
\begin{cases}
d_\infty, D(i,j) \leq n \\
\mathbf{M}_{ij},\, \text{otherwise},
\end{cases}
\end{equation}
where $D(i,j)$ is the smallest number of bonds to travel from atom $i$ to atom $j$ and $d_\infty$ is some number greater than the maximal  filtration value.


\subsubsection{Interactive persistent homology}
In protein-ligand binding analysis and/or mutation analysis, we are interested in the change of topological invariants induced by binding interactions and/or mutations.   Similar to the idea of multi-level persistent homology, we can design a distance matrix to focus on the interactions of interest. 
For a set of atoms, $A = A_1\cup A_2$ with $A_1\cap A_2 = \emptyset$ where only interactions between atoms from $A_1$ and atoms from $A_2$ are of interest \cite{ZXCang:2017b}. The interactive distance matrix $\mathbf{\widehat{M}}$ is defined as
\begin{equation}\label{eq:interactive}
\mathbf{\widehat{M}}_{ij} =
\begin{cases}
\mathbf{M}_{ij},\,\text{if}\, a_i\in A_1,\, a_j\in A_2 \, \text{or} \,a_i\in A_2,\,a_j\in A_1, \\
d_\infty,\, \text{otherwise},
\end{cases}
\end{equation}
where $\mathbf{M}$ is the original distance matrix induced from Euclidean metrics or other correlation function based distances, $a_i$ and $a_j$ are atoms $i$ and $j$ , and $d_\infty$ is a number greater than the maximal filtration value. In applications, $A_1$ and $A_2$ can be respectively sets of atoms of the protein and set of atoms of the ligand in a protein ligand complex. In this case, 
 the characterization of interactions between ligand and protein is an important task. In the modeling of site specific  mutation induced protein stability changes, $A_1$ could be the set of atoms at the mutation site and $A_2$ could be the set of atoms of  surrounding residues close to the mutation site. Similar treatment can be used for protein-protein and protein-nucleic acid interactions.

\subsubsection{Correlation function based persistent homology}

For biomolecules, the interaction strength between pair of atoms usually does not align linearly to their Euclidean distances. For example, van der Waals interaction is often described by the  Lennard-Jones potential.  Therefore, kernel function filtration can be used to emphasize certain geometric scales.  Correlation function based filtration matrix  was introduced in our earlier work   \cite{KLXia:2014c}:
\begin{equation}\label{eq:Correlmatrix}
\mathbf{\bar{M}}_{ij} = 1-\Phi(d_{ij},\eta_{ij}),
\end{equation}
where $\Phi(d_{ij},\eta_{ij})$ is a radial basis function and  $\eta_{ij}$ is a scale parameter. 
This filtration can be incorporated in the element specific persistent homology
\begin{equation}\label{eq:ESPHmatrix2}
\mathbf{\grave{M}}_{ij} = 
\begin{cases}
d_\infty,\,          \rm{if\, atom}\, $i$ \,\rm{or~ atom}\, $j$ \, \in {\cal U}  , \\
1-\Phi(d_{ij},\eta_{ij}),\, \rm{otherwise}.
\end{cases}
\end{equation}
Additionally, one can simultaneously use two or more correlation functions characterized by different scales  to generate a multiscale representation of biomolecules \cite{DDNguyen:2017d}.  

\paragraph{Flexibility and rigidity index based filtration matrix}

One form of the correlation function based filtration matrix is constructed by flexibility and rigidity index. In this case, the  
 Lorentz function is used in Eq. (\ref{eq:ESPHmatrix2})
\begin{equation}\label{eq:fri}
\Phi(d_{ij};\eta_{ij},\nu) = \frac{1}{1+\left(\frac{d_{ij}}{\eta_{ij}}\right)^\nu},
\end{equation}
where $d_{ij}$ is the Euclidean distance between point $i$ and point $j$ and $\eta_{ij}$ is a parameter controlling the scale and is related to radius of  two atoms. When distance matrices based on such correlation functions are used, patterns at different spatial scales can be addressed separately by altering the scale parameter $\eta_{ij}$.  Note that the rigidity index is given by \cite{KLXia:2013d }
\begin{equation}\label{eq:rigidity}
\mu_i=\sum_j\Phi(d_{ij};\eta_{ij},\nu). 
\end{equation}
This expression is closely related to the rigidity density based volumetric filtration \cite{KLXia:2015e}.
 
\subsubsection{Electrostatic  persistence}

Electrostatic effects are some of the most important effects in biomolecular structure, function and dynamics. The embedding of   electrostatics in topological invariants  is of particular interest and can be very useful in describing highly charged biomolecules such as nucleic acids and their complexes.  We  introduce electrostatics interaction induced distance functions in Eq. (\ref{eq:chg}) to address the electrostatic interactions among charged atoms. The abstract distance between two charged particles are rescaled according to their charges and their geometric distance, and is modeled as
\begin{equation}\label{eq:chg}
\Phi(d_{ij}, q_i, q_j; c) = \frac{1}{1+\exp(-cq_iq_j/d_{ij})},
\end{equation}
where $d_{ij}$ is the distance between the two atoms, $q_i$ and $q_j$ are the partial charges of the two atoms, and $c$ is a nonzero tunable parameter. $c$ is set to a positive number if opposite charge interactions are to be addressed and is set to a negative number if like charge interactions are of interest. The form of the function is adopted from sigmoid function which is widely used as an activation function in neural networks. Such function regularizes the input signal to the $[0,1]$ interval. Other functions can be similarly used. 
 This formulation can be extended to systems with dipole or higher order multipole approximations to electron density. The weak interactions due to long distances or neutral charges result in correlation values close to 0.5. When $c>0$, the repulsive interaction and attractive interaction deliver the correlation values in $(0.5,1)$ and $(0,0.5)$ respectively. The distances induced by $\Phi(d_{ij}, q_i, q_j; c)$ are used to characterize electrostatic effects. The parameter $c$ is rather physical but chosen to effectively spread the computed values over the $(0,1)$ interval so that the results can be used by machine learning methods. Another simple choice of charge correlation functions is $$\Phi(d_{ij}, \eta_{ij}, q_i, q_j) = q_iq_j\exp(-d_{ij}/\eta_{ij}).$$ However, this choice will lead to a different filtration domain.   
Additional, a charge density can be constructed 
\begin{equation}\label{eq:chargedensity}
\mu^c({\bf r}) =\sum_j  q_j \exp(-\|{\bf r}-{\bf r}_j\| /\eta_{j}),
\end{equation}
where ${\bf r}$ is a position vector, $\|{\bf r}-{\bf r}_j\|$ is the Euclidean distance between ${\bf r}$ and $j$th atom position ${\bf r}_j$ and $\eta_j$ is a scale parameter.  Equation (\ref{eq:chargedensity}) can be used for electrostatic  filtration as well. In this case, the filtration parameter can be the charge density value and cubical complex based filtration can be used.

\subsubsection{Multicomponent persistent homology}

Multicomponent persistent homology refers to the construction of multiple persistent homology components from a given object to describe its properties. Obviously, element specific persistent homology leads to multicomponent persistent homology. Nevertheless, in element specific persistent homology, the emphasis is given to the appropriate selection of important elements for describing certain biological properties or functions. For example, in biological context,  electronegative atoms are selected for describing hydrogen bond interactions, polar atoms are selected for describing hydrophilic  interactions, and   carbon atoms are selected for describing  hydrophobic interactions. Note that in chemical context, an atom may have many sharply different chemical and physical properties, depending on its oxidation states.  
Whereas, in multicomponent persistent homology, the emphasis is placed on the systematic generation of topological invariants from different combinatorial possibilities and the construction of 2D or  high-dimensional persistent maps for deep convolutional neural networks. Additionally, multicomponent persistent homology can also be constructed from  the combination of other persistences, such as electrostatic persistent homology, resolution induced persistent homology, etc.  

\subsection{Feature generation from topological invariants}\label{sec:SDB}

Barcode representation of topological invariants offers a visualization of persistent homology analysis.   In machine learning analysis, one needs to convert the barcode representation of topological invariants into machine learning feature vectors. To this end, we introduce two methods, i.e., counts in bins and barcode statistics,  to generate feature vectors from sets of barcodes. These methods are discussed below.

\paragraph{Counts in bins} 
For a given set of atoms $A$, we denote its barcodes as $\mathbf{B}=\{I_\alpha\}_{\alpha\in A}$  and  characterize each bar by an interval $I_\alpha=[b_\alpha,d_\alpha]$, where $b_\alpha$ and $d_\alpha$ are respectively the birth and death positions on the filtration axis. 
The length of each bar, or the persistence of topological invariant is given by  $p_\alpha=d_\alpha - b_\alpha$. 
To locate the position position of all bars and persistences, we further split the set of barcodes on the filtration axis into a predefined $N$ bins $\mathbf{Bin}=\{{rm Bin}_i\}_{i=1}^{N}$  where ${\rm Bin}_i = [l_i, r_i]$, where $l_i$ and $r_i$ are the left and the right positions of the $i$th bin. We generate features by counting the numbers of births, deaths, and persistences  in each bin, which leads to three counting feature vectors, namely, counts of birth  $F^C_b$, death $F^C_d$, and persistence $F^C_p$,  
\begin{equation}\label{eq:bincount}
\begin{aligned}
F^C_{b,i}(\mathbf{B}) &= \|\{[b_\alpha, d_\alpha]\in \mathbf{B} \vert l_i \leq b_\alpha \leq r_i\}\|, 1\leq i\leq N, \\
F^C_{d,i}(\mathbf{B}) &= \|\{[b_\alpha, d_\alpha]\in \mathbf{B} \vert l_i \leq d_\alpha \leq r_i\}\|, 1\leq i\leq N, \\
F^C_{p,i}(\mathbf{B}) &= \|\{[b_\alpha, d_\alpha]\in \mathbf{B} \vert b_\alpha\leq r_i\, {\rm or}\, l_i \leq d_\alpha\}\|, 1\leq i\leq N, \\
\end{aligned}
\end{equation}
where $\|\cdot\|$ is number of elements in a set. 
Note that the above discussion should be applied to three topological dimensions, i.e., barcodes of Betti-0 ($\mathbf{B^0}$), Betti-1 ($\mathbf{B^1}$) and Betti-2 ($\mathbf{B^2}$).  In general, this approach enables the description bond lengths, including the length of non-covalent interactions, in biomolecules and is referred as  binned persistent homology  in our earlier work \    \cite{ZXCang:2017a,ZXCang:2017b,ZXCang:2017c}.
 
\paragraph{Barcode statistics}
Another method of feature vector generation from a set of barcodes is  to extract important statistics of  barcode  collections such as maximum values and standard deviations. Given a set of bars $\mathbf{B}=\{[b_\alpha, d_\alpha]\}_{\alpha\in A}$, we define sets of $\mathbf{Birth} = \{b_\alpha\}_{\alpha\in A}$, $\mathbf{Death} = \{d_\alpha\}_{\alpha\in A}$, and $\mathbf{Persistence} = \{d_\alpha - b_\alpha\}_{\alpha\in A}$. Three statistic feature vectors $F^S_b$, $F^S_d$, and $F^S_p$ can then be generated in the sense of the statistics of the collection of barcodes. For example, $F^S_b$  consists of ${\rm avg}(\mathbf{Birth}), {\rm std}(\mathbf{Birth}), \max(\mathbf{Birth}), \min(\mathbf{Birth}), {\rm sum}(\mathbf{Birth})$, and ${\rm cnt}(\mathbf{Birth})$, where ${\rm avg}(\cdot)$ is the average value of a set of numbers, ${\rm std}(\cdot)$ is the standard deviation of a set of numbers, $\max(\cdot)$ and $\min(\cdot)$ are maximum and minimum values in a set of numbers, ${\rm sum}(\cdot)$ is the summation of elements in a set of numbers, and ${\rm cnt}(\cdot)$ is the count of elements in a set. The generation of $F^S_d$ is the same by examining the set $\mathbf{Death}$. $F^S_p$ contains the same information with two extra terms, the birth and death values of the longest bar. Statistic feature vectors are collected from barcodes of three topological dimensions, i.e., Betti-0, Betti-1 and Betti-2. \color[rgb]{0,0,0}

\paragraph{Persistence diagram slice and statistics}

\begin{figure}
\begin{center}
\includegraphics[scale=0.3]{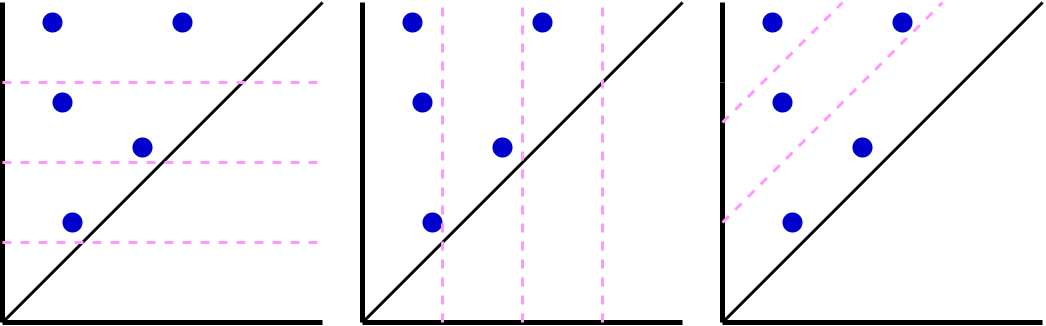}
\caption{Illustration of dividing set of barcodes into subsets. The barcodes are plotted as persistence diagrams with the horizontal axis being birth and the vertical axis being death. From left to right, the subsets are generated according to the slicing of death, birth, and persistence values.} \label{fig:pdbs}
\end{center}
\end{figure}

A more thorough description of sets of barcodes is to first divide the sets into subsets and extract features analogously to the \textit{barcode statistics} method. As shown in Figure \ref{fig:pdbs}, a persistence diagram can be divided into slices in different directions. The barcodes that fall in each slice form a subset. Each subset is described in terms of feature vector by using the \textit{barcode statistics} method. When the persistence diagram is sliced horizontally, members in each subset have similar death values and the barcode statistics feature vector is generated for the set of birth values. Similarly, members in each subset have similar birth values if the persistence diagram is sliced vertically, and the barcode statistics feature vector is generated for the set of death values. The barcode statistics feature vectors are generate for both set of birth values and set of death values if the persistence diagram is sliced diagonally, where members in each subset have similar persistence. This type of feature vector generation describes the set of barcodes in more detail but will produce long feature vectors.

\paragraph{2D representation}
The construction of multi-dimensional persistence is an interesting topic in persistent homology. In general, it is believed that multi-dimensional persistent has better representational power for complex systems described by multiple parameters \cite{Carlsson:2009theory}.  Although multidimensional persistence is hard to compute, one can compute persistence for one parameter while fixing the rest of the parameters to a sequence of fixed values. In the case where there are two parameters, a bifiltration can be done by taking turns to fix one parameter to a sequence of fixed values while computing persistence for the other parameter. For example, one can take a sequence of resolutions and compute persistence for distance with each fixed resolution. The sequence of outputs can be stacked to form a multidimensional representation \cite{KLXia:2015c}.  

Computing persistence multiple times and stacking the results is especially useful when the parameters that are not chosen to be the filtration parameter are naturally discrete with underlying orders. For example, the multicomponent or element specific persistent homology will result in many persistent homology computations over different selections of atoms. These results can be ordered by the percentage of atoms used of the whole molecule or by their importance in classical machine learning methods. 
Also, multiple underlying dimensions exist in the element specific persistent homology characterization of molecules. 
This property enables 2D or 3D topological representation of molecules. 
Based on the observation that the performance of the predictor degenerates when too many element combinations are used, we order the element combinations according to their individual performance on the task using methods of ensemble of trees. Combining the dimension of spatial scale and dimension of element combinations, a 2D topological representation is obtained. Such representation is expected to work better in the case of complex geometry such as protein-ligand complexes. With $\mathbf{E}=\{E_j\}_{j=1}^{N_E}$ denoting the collection of element combinations ordered by their individual performance on the task and $\mathbf{B}^{dim}(E_i)$ being the Betti-$dim$ barcodes obtained with atoms of element combination $E_j$, eight 2D representations are defined as 
\begin{equation}
\footnotesize
\{ F^C_{d,i}(\mathbf{B}^0(E_j)),F^C_{p,i}(\mathbf{B}^0(E_j)),F^C_{b ,i}(\mathbf{B}^1(E_j)),F^C_{d,i}(\mathbf{B}^1(E_j)),F^C_{p,i}(\mathbf{B}^1(E_j)),F^C_{b,i}(\mathbf{B}^2(E_j)),F^C_{d,i}(\mathbf{B}^2(E_j)),F^C_{p,i}(\mathbf{B}^2(E_j))\}_{i=1,\cdots, N}^{j=1,\cdots, N_E},
\end{equation}
where $F^C_{\gamma,i}$ with $\gamma=b, d, p$ is the barcode counting rule defined in Eq. (\ref{eq:bincount}). For Betti-0, since all bars start from zero, there is no need for $F^C_{b,i}(\mathbf{B}^0(E_j))$. These eight 2D representations are regarded as eight channels of a 2D topological image. In protein-ligand binding analysis, 2D topological features are generated for the barcodes of a protein-ligand complex and for the  differences between barcodes of the protein-ligand complex and those of the protein. Therefore, we have a total of  16 channels in a 2D image for  the protein-ligand complex. This 16-channel image can be fed into the training or the prediction of  convolutional neural networks.

\begin{figure}
\begin{center}
\includegraphics[scale=0.6]{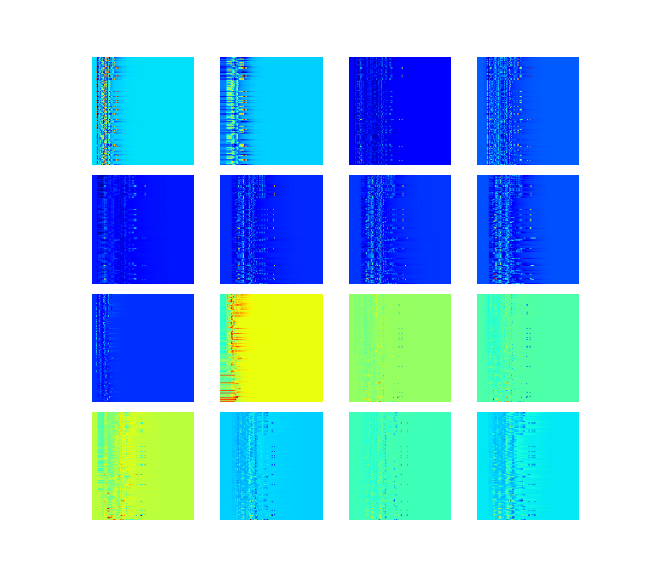}
\end{center}
\caption{The 2D topological maps of the 16 channels of sample 1wkm in PDBBind. The top 8 maps are for protein-ligand complex and the other 8 maps are for the difference between protein-ligand complex and protein only. For each map, the horizontal axis is the dimension of spatial scale and the vertical axis is element combinations ordered by their importance.}\label{fig:feature2d}
\end{figure}
 
In the characterization of protein-ligand complexes using alpha complexes, 2D features are generated from the alpha complex based on persistent homology computations of protein and protein-ligand complex.  A total of 128 element combinations are considered. The $[0,12]$\AA~ interval is divided into 120 equal length bins, which defines the resolution of  topological images. Therefore, the input feature for each sample is a 120$\times$128$\times$16 tensor.   Figure \ref{fig:feature2d} illustrates 16 channels of sample 1wkm in PDBBind database. These images are directly used in deep convolutional neural networks for training and prediction.

\subsection{Machine learning algorithms}

We discuss three machine learning  algorithms, including k-nearest neighbors (KNN) regression, gradient boosting trees and deep convolutional neural networks.

\subsubsection{K-nearest neighbors algorithm via barcode space metrics} 
One of the simplest machine learning algorithms is k-nearest neighbors (KNN) for classification or for regression. In KNN regression, for a given object, its property values is obtained by the average or the weighted average of the values of its $k$ nearest neighbors induced by the given metric. Then, the problem becomes how to construct a metric on the dataset. 

In the present work, instead of constructing feature vectors from sets of barcodes, the similarity between biomolecules can simply be derived from distances between sets of barcodes generated from different biomolecules. Popular barcode space metrics include the bottleneck distance induced metric \cite{cohen2005stability} and more generally, the Wasserstein metrics \cite{cohen2010lipschitz,carlsson2014topological}. The definition of the two metrics is summarized as follows. 

\paragraph{Bottleneck distance metric }
Given two bars $I_1 = [b_1,d_1]$ and $I_2 = [b_2,d_2]$ regarded as ordered pairs in ${\mathbb R}^2$, the $l^{\infty}$ distance between the two bars is defined as $\Delta (I_1,I_2) = \max(\vert b_2-b_1\vert, \vert d_2-d_1\vert)$. For a single bar $I=[b,d]$, $\lambda(I)$ is defined as $\lambda(I)=(d-b)/2$ which helps reflect the difference between the existence of the bar itself and the void. For two finite sets of barcodes $\mathbf{B}_1=\{I^1_{\alpha}\}_{\alpha\in A}$ and $\mathbf{B}_2=\{I^2_{\beta}\}_{\beta\in B}$ and a bijection $\theta$ from $A'\subseteq A$ to $B'\subseteq B$, the penalty of $\theta$ is defined as
\begin{equation}\label{eq:infpenalty}
P(\theta) = \max(\max\limits_{\alpha\in A'}(\Delta(I^1_{\alpha}, I^2_{\theta(\alpha)})), \max\limits_{\alpha\in A-A'}(\lambda(I^1_{\alpha})), \max\limits_{\beta\in B-B'}(\lambda(I^2_{\beta}))).
\end{equation}
Intuitively, a bijection $\theta$ is penalized for linking two bars with large difference and for ignoring long bars from either set. The bottleneck distance is defined as $d^{\infty}(I^1,I^2) = \min\limits_{\theta}P(\theta)$, where the minimum is taken over all possible bijections from subsets of $A$ to subsets of $B$. 

\paragraph{Wasserstein metric}
The Wasserstein metric, a $L_p$ generalized analog to the bottleneck distance can be defined with the penalty \cite{carlsson2014topological}
\begin{equation}\label{eq:ppenalty}
P^p(\theta) = \sum\limits_{\alpha\in A'}\Delta(I^1_{\alpha},I^2_{\theta(\alpha)})^p  + \sum\limits_{\alpha\in A-A'}\lambda(I^1_{\alpha})^p + \sum\limits_{\beta\in B-B'}\lambda(I^2_{\beta})^p
\end{equation}
and the corresponding distance $d^p(\mathbf{B}_1,\mathbf{B}_2) = (\min\limits_{\theta}P^p(\theta))^\frac{1}{p}$. It reduces to the bottleneck distance by setting $p=\infty$. In this work, we choose $p=2$. 

Wasserstein metric measures the closeness of barcodes generated from different biomolecules.  It will be interesting to consider other distances for metric spaces, such as Hausdorff distance,  Gromov-Hausdorff distance \cite{burago2001course}, and Yau-Hausdorff distance \cite{tian2015two} for biomolecular   analysis. However, an  exhaustive study of this issue is beyond the scope of the present work.   

The barcode space metrics can be directly used to assess the representation power of various persistent homology methods on biomolecules without potential overfitting effects induced by manually generated feature vectors. We show in the section of results  that the barcode space metrics induced similarity measurement is significantly correlated to molecule functions. 

Wasserstein metric measures from biomolecules can also be directly implemented in a kernel based method such as nonlinear support vector machine algorithm for classification and regression tasks. However, this aspect is not explored in the present work.

\subsubsection{Gradient boosting trees}

Gradient boosting trees is an ensemble method which ensembles individual decision trees to achieve the capability of learning complex feature target maps and can effectively prevent overfitting by using shrinkage technique. The gradient boosting trees method is implemented using the GradientBoostingRegressor module in scikit-learn software package \cite{scikit-learn}. A set of parameters found to be efficient in our previous study on the protein-ligand binding affinity prediction  \cite{ZXCang:2017b} is used uniformly unless specified. The parameters used are \textit{n\_estimators=20000, max\_depth=8, learning\_rate=0.005, loss='ls', subsample=0.7, max\_features='sqrt'.}

\begin{figure}
\begin{center}
\includegraphics[scale=0.2]{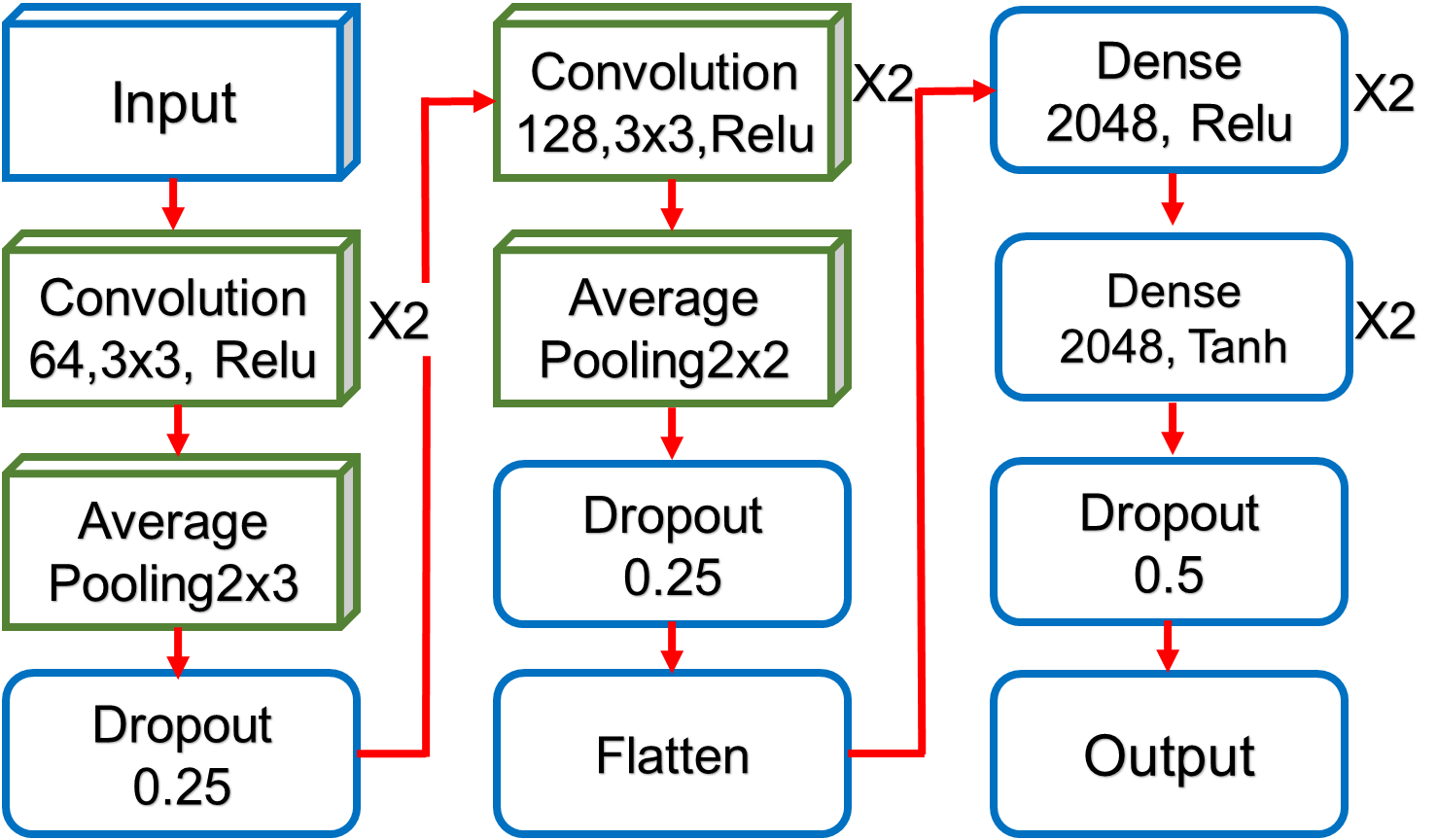}
\caption{The architecture of the deep  convolutional neural network. The structured layers are shown in boxes and the unstructured layers are shown in rectangles. The convolution layers are drawn with number of filters, filter size, and activation function. The dense layers are drawn with number of neurons and activation function. The pooling size of the pooling layers and dropout rate of the dropout layers are listed. The layers that are repeated twice are marked with ``$\times$2" sign on the right side of the layer.}\label{fig:convnet}
\end{center}
\end{figure}

\subsubsection{Deep convolutional neural networks}
A widely used convolutional neural network architecture is employed beginning with convolution layers followed by dense layers. Due to the limited computation resources, parameter optimization is not performed, while most parameters are adopted from our earlier work \cite{ZXCang:2017c}. Reasonable parameters are assigned manually. The detailed architecture is shown in Figure \ref{fig:convnet}. The Adam optimizer with learning rate 0.0001 is used. The loss function is the mean squared error function. The network is trained with a batch size of 32 and 500 epochs. The training data is shuffled for each epoch. The deep convolutional neural network is implemented using Keras \cite{chollet2015keras} with Theano backend \cite{2016arXiv160502688short}.

\section{Results and discussion}\label{Results}

 Rational drug design and  discovery have rapidly evolved into some of the most important and exciting research fields  in   medicine and biology. These approaches potentially have a profound impact on human health. The ultimate goal is to determine and predict whether a given drug candidate will bind to a target so as to activate or inhibit its function, which   results in a therapeutic benefit to the patient. 
Virtual screening is an important process in rational drug design and discovery which aims to identify actives of a given target from a library of small molecules. There are mainly two types of screening techniques, ligand-based and structure-based. Ligand-based approach depends on the similarity among small molecule candidates. Structure-based approach trys to dock a candidate molecule to the target protein and judge the candidate with the modeled binding affinity based on the docking poses. In structure-based screening, knowledge based rescoring  methods using machine learning or deep learning approaches have shown improvements when applied on top of docking algorithms \cite{pereira2016boosting, durrant2013comparing, arciniega2014improvement}. We also apply our method as a rescoring machine to rerank the candidates based on docking poses generated by docking software. 

\begin{figure}
\begin{center}
\includegraphics[scale=0.45]{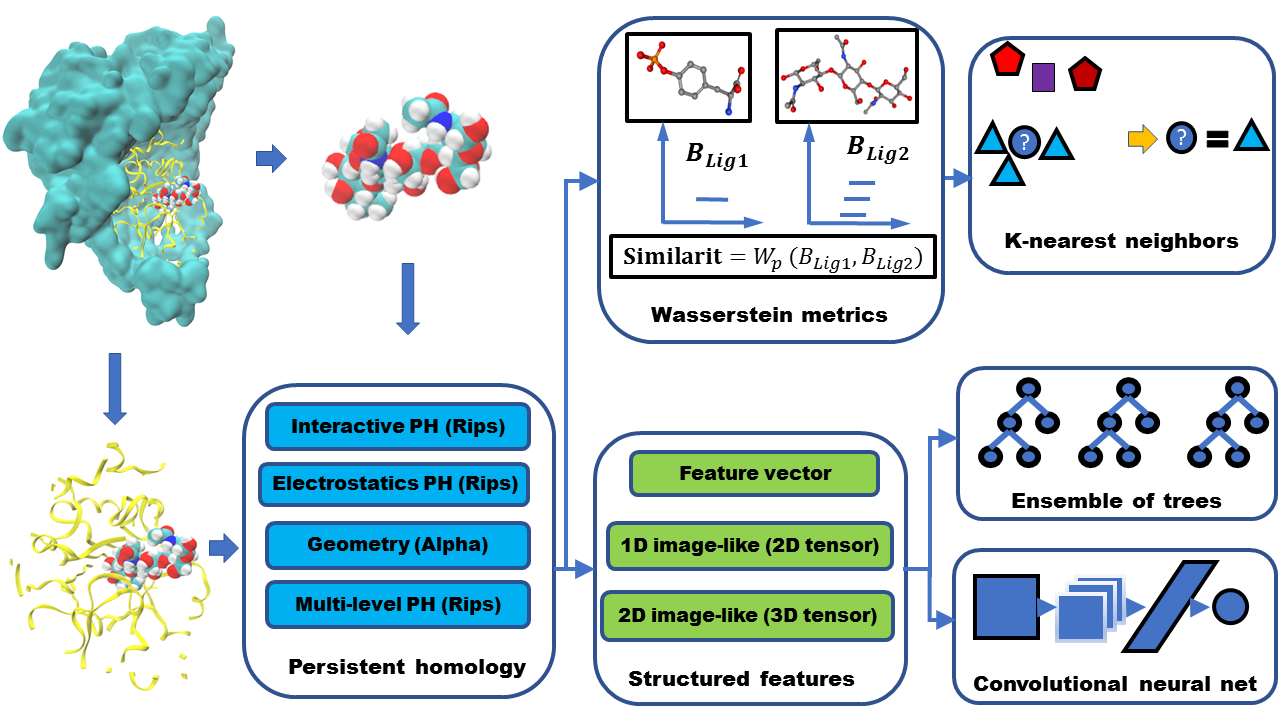}
\caption{An illustration of the topology based machine learning algorithms used scoring and virtual screening.}\label{fig:illustration}
\end{center}
\end{figure}

This section explores the representational power of the proposed persistent homology methods for the prediction of protein-ligand binding affinities  and the discrimination of actives and non-actives for protein targets.  To this end, we use the present method to investigate  three types of problems. First,  we consider a ligand based protein-ligand binding affinity predictions. This problem is designed to examine the representability of the proposed topological methods for small molecules.   Additionally, we study protein-ligand complex based binding affinity prediction. This problem enables us to understand the capability of the proposed topological methods for dealing with protein-ligand complexes. Finally, we examine the complex based classification of active ligands and non-active decoys, i.e., structure-based virtual screening (VS). The optimal selection of features and methods are determined by studying the first two applications and this leads to the main application studied in this work, the topological structure-based virtual screening. Computational algorithms used in this study are illustrated in Fig. \ref{fig:illustration}. 

\subsection{Ligand based protein-ligand binding affinity prediction} \label{Exp:LigandBased}

We consider the protein-ligand binding affinity prediction using a ligand based approach. Essential hypothesis is that, for a given binding site on a target protein, similar ligands or ligands with similar decisive substructures have similar binding free energies. One data set and two machine learning algorithms, i.e., gradient boosting trees and KNN with Wasserstein metrics are employed in this study to analyze the representational power of the proposed method.  Cross-validations within the ligand set of each protein target are carried out.

\begin{figure}
\begin{center}
\includegraphics[scale=0.25]{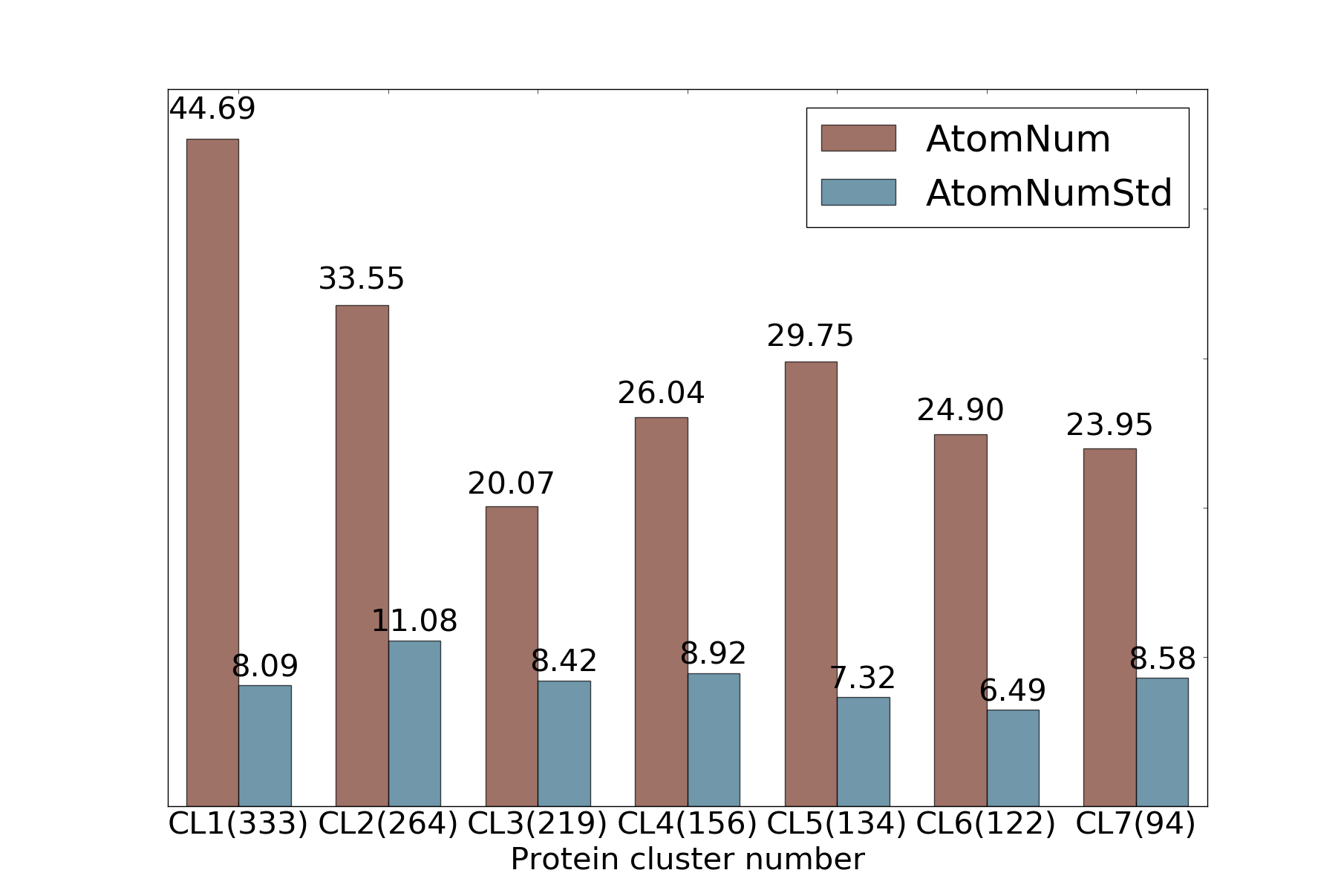}
\end{center}
\caption{Statistics of ligands in 7 protein clusters in S1322. The average numbers of heavy atoms of a ligand in each protein cluster are shown in red and the standard deviations of number of heavy atoms across each protein cluster are shown in blue.
The number of ligands in each cluster is given in parentheses. 
}\label{fig:ligstat}
\end{figure}

\paragraph{Data set}
To assess the representational ability of the present persistent homology algorithms on small molecules, we use a high quality data set of 1322 protein-ligand complexes with binding affinity data involving 7 protein clusters introduced earlier (denoted as S1322) \cite{BaoWang:2016FFTB}. It is a subset of the PDBBind v2015 refined set and its detail   is given in the  Supporting Material of Ref.  \cite{BaoWang:2016FFTB}.  We consider a ligand based approach to predict the binding affinities of protein-ligand complexes in various protein clusters. As such, only the ligand information is used in our topological analysis. The ligand structures are taken from PDBBind database without modification. Numbers of ligands in  protein clusters range from 94 to 333. Statistics of  the data set in terms of the average  number  of heavy atoms and its  standard deviations is given in  Fig. \ref{fig:ligstat}. In this dataset, cluster 1 has a relatively large binding site that is able to accommodate large ligands, while cluster 3 has a relatively small binding site.

\paragraph{Feature vectors for gradient boosting trees}
In this test, Rips complex based and alpha complex based persistent homology computations up to Betti-2 are performed for a variety of atom collections with different element types using the Euclidean metric and multi-level distance defined in Eq. \ref{eq:order1matrix}. Two types of features are generated and are denoted by $F^C$, which is a combination of $F^C_b$, $F^C_d$, and $F^C_p$, and $F^S$, which is a combination of $F^S_b$, $F^S_d$, and $F^S_p$. For sets of Betti-0 bars, only $F^C_d$ and $F^S_d$ are computed. In each protein cluster, 10-fold or 5-fold cross validation is repeated 20 times for each subset of feature vectors depending on selected element type.  The median Pearson correlation coefficients and the root-mean-square error (RMSE) in kcal/mol are reported. For Rips complex, both level 0 computation with distance matrix $\mathbf{M}$ and level 1 computation with  distance matrix $\mathbf{\widetilde{M}}^1$ as defined in Eq. (\ref{eq:ordernmatrix}) are performed. A comparison of these results  is shown in Table S2. 
The results corresponding to alpha complex are shown in Table S1. \%ref{tab:alpha}. 
The average performance for alpha complex and Rips complex has a Pearson correlation coefficient of 0.987.

\begin{figure}
\begin{center}
\includegraphics[scale=0.4]{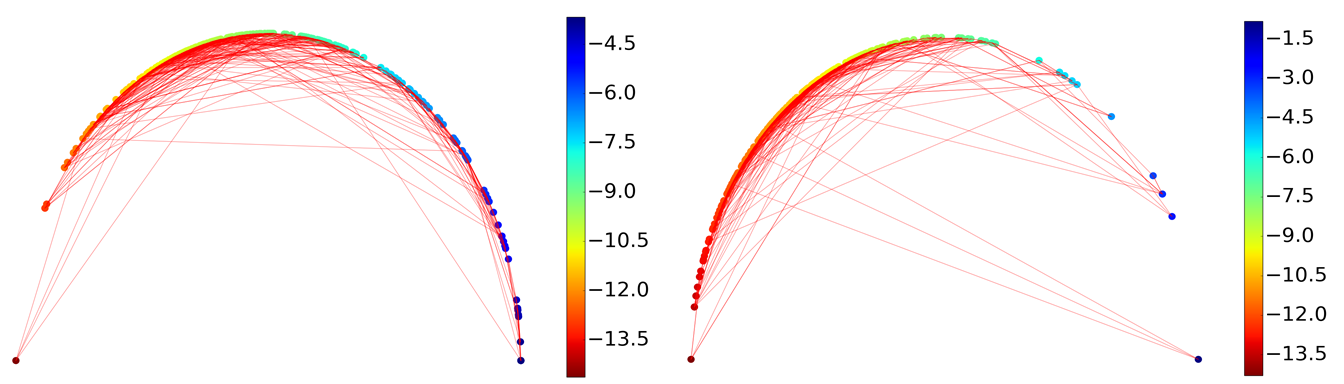}
\end{center}
\caption{An illustration of similarities between ligands measured by their  barcode space Wasserstein distances. Ligands are ordered according to their binding affinities and are represented as dots on the semicircle. Each dot is connected with two nearest neighbors based on their barcode space  Wasserstein distances. An optimal prediction would be achieved if lines stay close to the semicircle. 
The majority of the connections stay near the boundary to the upper half sphere demonstrating that barcode space metric based  Wasserstein distance measurement reflects the similarity in function, i.e., the binding affinity in this case. The protein clusters with the best and the worst performance are shown. (a) Protein cluster 2. (b) Protein cluster 3.}
\label{fig:KNN}
\end{figure}

\paragraph{Barcode space metrics for k-nearest neighbor regression}
The barcodes generated using Rips complex with distance matrices $\mathbf{M}$ and $\mathbf{\widetilde{M}}^1$ are collected and the distance between each pair of sets of barcodes are measured using the Wasserstein metric $d^2$. Leave-one-out prediction for every sample is performed with k-nearest neighbor regression with $k=3$ within each protein cluster based on the Wasserstein metric. The results are shown in Table S3. 
The performance of the best performing and the worst performing protein clusters is shown in Figure \ref{fig:KNN}. The better the performance, the closer the lines are to the semicircle. 

The experiments done for this section are summarized in Table \ref{tab:ligandbasedaffinity}.

\begin{table}[h]
\footnotesize
\begin{center}
\begin{tabular}{|l|p{12cm}|} \hline
Experiment & Description \\ \hline
A-B012-E-C-GBT & The barcodes are generated using alpha complex on different sets of atoms based on different element combinations. The features are constructed using Betti-0, Betti-1, and Betti-2 barcodes following the \textit{counts in bins} method with bins equally dividing the interval $[0,5]$. Here 32 different element combinations are considered, including \{C, N, O, S, CN, CO, CS, NO, NS, OS, CNO, CNS, COS, NOS, CNOS, CNOSPFClBrI, H, CH, NH, OH, SH, CNH, COH, CSH, NOH, NSH, OSH, CNOH, CNSH, COSH, NOSH, CNOSH, CNOSPFClBrIH \}. Gradient boosting trees (GBTs) with the structured feature matrix are used for this computation. \\ \hline
A-B012-E-S-GBT & The barcodes same as those used in A-B012-E-C-GBT are used. Instead of \textit{ counts in bins }, the \textit{Barcode statistics} method is used to generate features. \\ \hline
A-B012-E-SS-GBT & The barcodes same as those used in A-B012-E-C-GBT are used. The \textit{persistence diagram slice and statistics} method is used to generate features. A uniform set of bins by dividing the interval [0,5] into 10 equal length bins is used to slice birth, death, and persistence values. \\ \hline
R-B012-E-S-GBT & Barcodes are generated using Rips complex with Euclidean distances. The features are generated following the \textit{barcode statistics} method. Here 36 element combinations are considered, i.e., \{ C, N, O, S, CN, CO, CS, NO, NS, OS, CNO, CNS, COS, NOS, CNOS, CNOSPFClBrI, H, CH, NH, OH, SH, CNH, COH, CSH, NOH, NSH, OSH, CNOH, CNSH, COSH, NOSH, CNOSH, CNOSPFClBrIH, CCl, CClH, CBr, CBrH \}. \\ \hline
R-B012-M1-S-GBT & The result is obtained with the same setup as R-B012-E-S-GBT except that the first level enrichment distance matrix $\mathbf{\widetilde{M}}^1$ is used instead of Euclidean distance. \\ \hline
R-B$n$-E-KNN & The Betti-$n$ barcodes  from Rips complex computation with Euclidean distance are used. K-nearest neighbor (KNN) regression is performed with Wasserstein metric $d^2$.  The leave-one-out validation is performed individually with each element combination and the average prediction of these element combinations is taken as the output result. The element combinations considered are \{ CNOS, CNOSPFClBrI, NOH, CNO, CNOSPFClBrIH\}. These combinations are selected based on their performance in the gradient boosting trees experiments. \\ \hline
R-B$n$-M1-KNN & The result is obtained with the same setup as R-B$n$-E-KNN except that the distance matrix $\mathbf{\widetilde{M}}^n$ is used instead of Euclidean distance. \\ \hline
\end{tabular}
\caption{Experiments for ligand-based protein-ligand binding affinity prediction of 7 protein clusters and 1322 protein-ligand complexes.}\label{tab:ligandbasedaffinity}
\end{center}
\end{table}

\begin{table}[ht]
\footnotesize
\begin{center}
\rowcolors{2}{gray!18}{white}
\begin{tabular}{llllllllll}
\toprule
\rowcolor{gray!28}
ID & Experiments & CL 1 (333) & CL 2 (264) & CL 3 (219) & CL 4 (156) & CL 5 (134) & CL 6 (122) & CL 7 (94) & Average \\
1 & A-B012-E-C-GBT & 0.695(1.63) & 0.836(1.18) & 0.690(1.52) & 0.642(1.38) & {\bf 0.840(1.30)} & 0.647(1.65) & 0.730(1.27) & 0.726(1.42) \\
2 & A-B012-E-S-GBT & 0.695(1.63) & 0.845(1.14) & 0.678(1.54) & {\bf 0.692(1.31)} & 0.828(1.35) & 0.702(1.54) & 0.739(1.25) & 0.740(1.39) \\
3 & A-B012-E-SS-GBT & 0.704(1.62) & 0.846(1.15) & 0.681(1.53) & 0.668(1.35) & 0.834(1.34) & 0.715(1.53) & 0.741(1.25) & 0.741(1.40) \\
4 & R-B012-E-S-GBT & 0.712(1.60) & 0.837(1.17) & 0.659(1.57) & 0.683(1.32) & 0.808(1.41) & 0.635(1.67) & {\bf 0.757(1.22)} & 0.727(1.42) \\
5 & R-B012-M1-S-GBT & \textbf{0.716(1.59)} & 0.836(1.17) & {\bf 0.706(1.48)} & 0.672(1.34) & 0.822(1.37) & {\bf 0.708(1.53)} & 0.746(1.24) & 0.744(1.39) \\
6 & 2+5 & 0.714(1.59) & {\bf 0.848(1.13)} & 0.699(1.50) & {\bf 0.692(1.31)} & 0.831(1.34) & {\bf 0.717(1.52)} & 0.747(1.24) & {\bf 0.750(1.38)} \\
7 & R-B0-E-KNN & 0.648(1.73) & 0.761(1.39) & 0.544(1.76) & 0.616(1.42) & 0.700(1.70) & 0.487(1.89) & 0.641(1.43) & 0.628(1.62) \\
8 & R-B1-E-KNN & 0.547(1.91) & 0.684(1.55) & 0.444(1.88) & 0.536(1.52) & 0.535(2.01) & 0.634(1.67) & 0.649(1.42) & 0.576(1.71) \\
9 & R-B2-E-KNN & 0.474(2.01) & 0.494(1.87) & 0.202(2.14) & 0.298(1.79) & 0.126(2.49) & 0.331(2.09) & 0.609(1.47) & 0.362(1.98) \\
10 & R-B0-M1-KNN & 0.581(1.85) & 0.771(1.35) & 0.516(1.80) & 0.601(1.44) & 0.672(1.76) & 0.485(1.90) & 0.644(1.43) & 0.610(1.65) \\
11 & R-B1-M1-KNN & 0.663(1.70) & 0.784(1.33) & 0.652(1.59) & 0.555(1.50) & 0.786(1.49) & 0.610(1.71) & 0.731(1.30) & 0.683(1.52) \\
12 & R-B2-M1-KNN & 0.675(1.67) & 0.803(1.28) & 0.577(1.72) & 0.531(1.52) & 0.655(1.81) & 0.617(1.72) & 0.648(1.42) & 0.644(1.59) \\
13 & Cons(7+8+9+10+11+12) & 0.698(1.66) & 0.817(1.28) & 0.620(1.68) & 0.645(1.41) & 0.756(1.68) & 0.658(1.68) & 0.739(1.31) & 0.705(1.49) \\
14 & 2+5 (5-fold) & 0.713(1.60) & 0.843(1.15) & 0.693(1.51) & 0.670(1.35) & 0.831(1.34) & 0.698(1.56) & 0.737(1.26) & 0.741(1.40) \\
15  &  FFT-BP (5-fold) \cite{BaoWang:2016FFTB} &  (1.93) &      (1.32) &       (2.01) &    (1.61) &       (2.02) &     (2.06) &     (1.71) & (1.81)\\
\bottomrule
\end{tabular}
\end{center}
\caption{ Pearson correlation coefficients with RMSE (kcal/mol) in parentheses for binding affinity predictions on  7 protein clusters (CL) in S1322. On the title row,  the numbers in parentheses denote the numbers of ligands in the cluster. The median results of 20 repeated runs are reported for the ensemble of trees based methods to account for randomness in the algorithm. For experimental labels, the first letter indicates the complex definition used, `A' for alpha complex and `R' for Rips complex. The second part starting with `B' followed by the integers indicates the Betti number used. The third part indicates the distance function used, `E' for Euclidean and `M1' for $\mathbf{\widetilde{M}}^1$. For row 1 through 5, the forth part shows the way of feature construction, `C' for counts in bins  and `S' for barcode statistics. The last part indicates the regression technique used, `GBT' for gradient boosting trees and 'KNN' for k-nearest neighbors. The detailed descriptions of the experiments are given in Table \ref{tab:ligandbasedaffinity}. Row 6 is the results using features of both row 2 and row 5. Row 13 is the consensus  results by taking the average of the predictions by row 7 through row 12. Except for specified, all results are obtained from 10-fold cross validations. FFT-BP 5-fold cross validation results were adopted from Ref.  \cite{BaoWang:2016FFTB}, where Pearson correlation coefficients were not given. 
}
\label{tab:lbfp}
\end{table}

\paragraph{Performance of multicomponent persistent homology }

It can be noticed from Table \ref{tab:lbfp} that topological features generated from barcode  statistics typically outperform  those created from  counts in bins.   R-B012-E-S-GBT and R-B012-M1-S-GBT perform similarly in the majority of the protein clusters whilst R-B012-M1-S-GBT which is based on $\mathbf{\widetilde{M}}^1$ significantly outperforms R-B012-E-S-GBT which is based on Euclidean distance in protein cluster 3 and 6. To assess in what circumstances does the multi-level persistent homology improve the original persistent homology characterization of small molecules, we analyze the statistics of the size of ligands in Figure \ref{fig:ligstat}. It turns out that protein cluster 3 has the smallest average number of heavy atoms and protein cluster 6 has the smallest standard deviation of the number of heavy atoms. This observation partially answers the question that in the cases where the small molecules are relatively simple and are relatively of similar size, multi-level persistent homology is able to enrich the characterization of the small molecules which further improves the robustness of the model. Such enrichment or improvement over the original persistent homology approach is mainly realized in higher dimensional Betti numbers, i.e. Betti-1 and Betti-2. In Table \ref{tab:lbfp}, the results with ID through 7 to 12 confirm that the Betti-0 features from computation with $\mathbf{\widetilde{M}}^1$ are inferior to the results with Euclidean distance whilst the Betti-1 or Betti-2 features based on $\mathbf{\widetilde{M}}^1$ outperforms the best result with Euclidean distance in most cases.

It is interesting to note that although  Wasserstein metric based KNN  methods are not as accurate as GBT approaches,  
the consensus result obtained by averaging over various   Wasserstein metric predictions  is quite accurate. Unlike GBT approaches,   Wasserstein metric utilizes only one type of barcodes.  In fact, R-B$0$-E-KNN,    R-B$1$-M1-KNN and  R-B$2$-M1-KNN work very well. 

Finally,  FFT-BP 5-fold cross validation results were obtained based on multiple  additive  regression  trees   and a set of physical descriptors, including geometry charge, electrostatics and van der Waals interactions for S1322 \cite{BaoWang:2016FFTB}. Since multiple  additive  regression  trees are  essentially the same as the GBT used in the present work, it is appropriate to compare the FFT-BP results with the GBT results in this work. The current topological descriptors built on only ligands have more predictive power than the physical descriptors built on protein-ligand complexes constructed in our earlier work  \cite{BaoWang:2016FFTB}.

\paragraph{Robustness of topological  models}

\begin{figure}
\begin{center}
\includegraphics[scale=0.35]{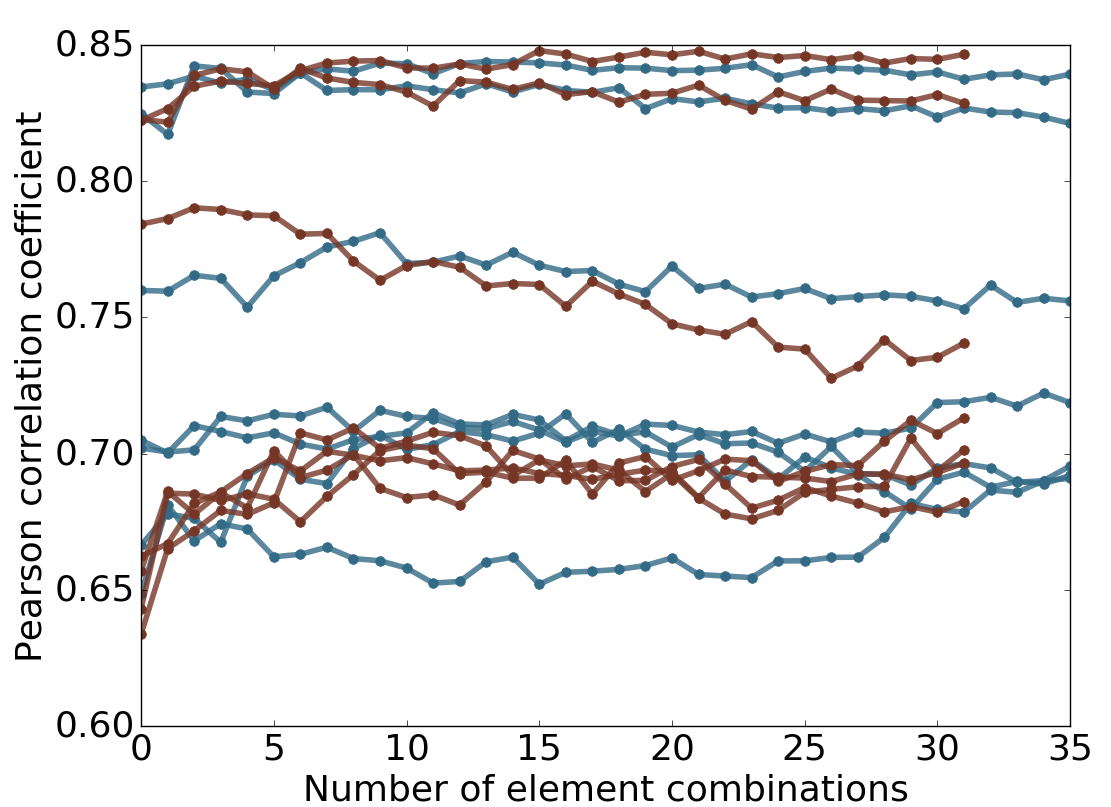}
\caption{The model performance against the number of element combinations involved in feature construction for 7 protein clusters in S1322. The results related to alpha complex are marked in red and Rips complex in blue.}
\label{fig:ligfeat}
\end{center}
\end{figure}

Certain elements such as Br are very rare in the data sets studied in this work. Considering only the elements of high occurrence will not hurt the performance on the validations performed. However, omitting the low occurrence elements will sacrifice the capability of the model to handle new data in which such elements play an important role. Therefore, we decide to keep the rare elements that  result in a large number of features and redundancy in features. For example, the element combinations CBrH and CH will probably deliver the same performance  for most of the samples in the data sets studied in this work. To test whether this redundancy causes degenerated results of the model, the features of one element combination is added to the model at a step and the model is validated with an accumulation of the added features at each step. The performance of the model is measured with Pearson correlation coefficient and is plotted against number of element combinations involved in Fig. \ref{fig:ligfeat}. For most cases in Fig.  \ref{fig:ligfeat}, the model is robust against the inclusion of more element combinations.


\subsection{Complex based protein-ligand binding affinity prediction}\label{Exp:ComplexBased}
 
Having demonstrated the representational power of the present topological method for characterizing small molecules, we further apply the proposed topological method to characterize both the protein and the ligand in a protein-ligand complex.  Biologically, we consider the same task, i.e., the prediction of protein-ligand binding affinity, with a different approach that is based on the structural information of both the protein and the ligand in the protein-ligand complex.  Only GBT and deep convolutional neural network algorithms are used in this section.  
 
\paragraph{Data sets}

\begin{table}[ht]
\begin{center}
\rowcolors{2}{gray!18}{white}
\begin{tabular}{lcccc}
\toprule
\rowcolor{gray!28}
Version &  Refined set & Training set& Core set (test set) &  Protein families \\
\midrule
v2007 & 1300 & 1105 & 195 & 65\\
v2013 & 2959 & 2764 & 195 & 65\\
v2015 & 3706 & 3511 & 195 & 65\\
v2016 & 4057 & 3767 & 290 & 58\\
\bottomrule
\end{tabular}
\end{center}
\caption{Number of complexes or number of protein families in  PDBBind data sets used in the present binding affinity prediction. 
Here training sets are obtained from corresponding refined sets, excluding the complexes in corresponding test sets. Protein families refer to those in the corresponding tests.  
 }\label{tab:Datasets}
\end{table}

The PDBBind database provides a comprehensive collection of structures of protein-ligand complexes and their binding affinity data  \cite{RenxiaoWang:2009Compare,PDBBind:2015}. The original experimental data in \href{www.rcsb.org}{Protein Data Bank (PDB)} \cite{Berman:2000} are selected  to PDBBind database based on certain quality  requirements and curated for applications. As shown in Table \ref{tab:Datasets}, this database is expanding on a yearly basis. It has become a standard resource for benchmarking computational methods and algorithms for protein-ligand binding analysis and drug design. Popular data sets include  version 2007 (v2007), v2013, and v2015. Among them, v2013 core set and v2015 core set are identical. A large number of scoring functions has been tested on these data sets. The latest version, v2016, has an enlarged core set, which contains 290 protein-ligand complexes to represent 58 protein families. Therefore, this test set should be relatively easier than v2015 core set, whose 195 complexes  involve  65 protein families. 

In the present machine learning study, we use four PDBBind core sets as our test sets. For each test set, the corresponding refined set, excluding the core set, is used as the training set. 

\paragraph{Groups of topological features and their performance in association with GBT  }

The experiments of protein-ligand-complex-based protein-ligand binding affinity prediction for the PDBBind datasets are summarized in Table \ref{tab:complexbasedaffinity}.

\begin{table}[h]
\footnotesize
\begin{center}
\begin{tabular}{|l|p{13cm}|} \hline
Experiment & Description \\ \hline
R-B0-I-C & Betti-0 barcodes from Rips complex computation with interactive distance matrix based on Euclidean distance are used. Features are generated  following \textit{counts in bins } method with bins $\{[0,2.5),[2.5,3),[3,3.5),[3.5,4.5),[4.5,6),[6,12]\}$. Element combinations used are all possible paired choices of one item from 
 \{ C, N, O, S, CN, CO, NO, CNO \} in protein and another item from 
\{ C, N, O, S, P, F, Cl, Br, I, CN, CO, CS, NO, NS, OS, CNO, CNS, COS, NOS, CNOS \} in ligand, which  result in a total of 160 combinations. \\ \hline
R-B0-I-BP & The persistent homology computation and feature generation is the same as R-B0-I-C. However, the element combinations used are all possible paired choices of one item from  \{ C, N, O, S \} in protein and   another item from 
\{ C, N, O, S, P, F, Cl, Br, I \} in ligand, which result in a total of 36 element combinations. \\ \hline
R-B0-CI-C & Betti-0 barcodes from Rips complex computation with interactive distance matrix based on the electrostatics correlation function defined in Eq. (\ref{eq:chg}) with the parameter $c=100$. The features are generated following \textit{counts in bins } method with bins $\{(0,0.1],(0.1,0.2],(0.2,0.3],(0.3,0.4],(0.4,0.5],(0.5,0.6],(0.6,0.7],(0.7,0.8],(0.8,0.9],(0.9,1.0)\}$. The element combinations used are all possible paired choices of one item from  \{ C, N, O, S, H \} in protein and another item from 
\{ C, N, O, S, P, F, Cl, Br, I, H \} in ligand, which result in a total of 50 element combinations. \\ \hline
R-B0-CI-B-S & The barcodes and element combinations are the same as those of R-B0-CI-B-C. The features are generated following the \textit{barcode statistics} method. \\ \hline
A-B12-E-S & Betti-1 and Betti-2 barcodes from alpha complex computation with Euclidean distance  are used. The element combinations considered are all heavy atoms and all carbon atoms. Features are generated following the \textit{barcode statistics} method. \\ \hline
\end{tabular}
\caption{Experiments for protein-ligand-complex-based protein-ligand binding affinity prediction for the PDBBind datasets.}\label{tab:complexbasedaffinity}
\end{center}
\end{table}

\begin{table}[ht]
\begin{center}
\rowcolors{2}{gray!18}{white}
\begin{tabular}{lllllll}
\toprule
\rowcolor{gray!28}
ID & Experiments & v2007 & v2013 & v2015 & v2016 & Average \\
\midrule
1 & R-B0-I-C  & 0.799 (2.01) & 0.741 (2.14) & 0.750 (2.11) & 0.813 (1.82) & 0.776 (2.02) \\
2 & R-B0-I-BP &  0.816 {\bf (1.94)} & 0.741 (2.13) & 0.750 (2.10) & 0.825 (1.78) & 0.783 (1.99) \\
3 & R-B0-CI-C & 0.791 (2.05) & 0.759 (2.10) & 0.738 (2.13) & 0.801 (1.87) & 0.772 (2.04) \\
4 & R-B0-CI-S & 0.773 (2.10) & 0.762 (2.12) & 0.749 (2.13) & 0.810 (1.86) & 0.774 (2.05) \\
5 & A-B12-E-S & 0.736 (2.25) & 0.709 (2.26) & 0.695 (2.27) & 0.752 (2.02) & 0.723 (2.20) \\
6 & 1+4       & 0.815 (1.95) & 0.780 (2.04) & 0.774 (2.04) & 0.833 (1.76)  & 0.801 (1.95) \\
7 & 2+4       & 0.806 (1.99) & 0.787 (2.04) & 0.770 (2.06) &  0.834 (1.77) & 0.799 (1.97) \\
8 & 1+4+5     & 0.810 (1.98) & 0.792 (2.02) & 0.786 (2.02) & 0.831 (1.76)  & 0.805 (1.95) \\
9 & 2+4+5     & 0.802 (2.01) & {\bf 0.796} (2.02) & 0.782 (2.04) & 0.822 (1.79) & 0.801 (1.97) \\
10 & 2D-CNN-Alpha   & 0.787 (2.02) & 0.781 ({\bf 1.98}) & 0.785 (1.95) & 0.837 (1.68) & 0.798 (1.91) \\
11 & 1D2D-CNN & 0.806 (1.95) & 0.781 ({\bf 1.98}) &{\bf  0.799 (1.91)} & {\bf 0.848 (1.64)} & 0.809 (1.87) \\ 
12 & RF::VinaElem$^a$ & 0.803 ({\bf 1.94}) \cite{HLi:2015} &  0.752 (2.03) \cite{HongJianLi:2015} &  - &  - & - \\
13 & RI-Score\cite{DDNguyen:2017d} $^b$ & {\bf 0.825}$^c$ (1.99) & - & 0.782$^d$ (2.05) & 0.815 (1.85) & -\\
\bottomrule
\end{tabular}
\end{center}
\caption{Pearson correlation coefficients with RMSE (kcal/mol) in parentheses for  predictions by various  groups of features on the four PDBBind core sets. Results of ensemble of trees based methods (rows 1 through 9) are the {\it median values} of 50 repeated runs to account for randomness in the algorithm. For the deep learning based methods (row 10 and 11), 100 independent models are generated in the first place. A consensus model is built by randomly choosing 50 models out of the 100, and the this process is repeated 1000 times with the median reported. 
The first letter indicates the definition of complex, `A' for alpha complex and `R' for Rips complex. The second part indicates the Betti numbers used. The third part indicates the distance function used, `I' for $\mathbf{\widehat{M}}_{ij}$ defined in \ref{eq:interactive}, `CI' for the one defined in \ref{eq:chg}, and `E' for Euclidean. The last part shows the way of feature construction, `C' for counts in bins, `S' for barcode statistics, and `BP' for only pair of two single elements.  The results reported in row 6 through 9 are obtained by combining the features of the rows with the corresponding numbers. $^a$ The authors did mot specify the number of repeated experiments and if the reported performance is the best or the median of the experiments. $^b$ The best results of the repeated runs with randomness are reported and the parameters in feature generation are optimized for each dataset. $^c$ The median PCC reported is 0.803. $^d$ The median PCC reported is 0.762.}
\label{tab:bfp}
\end{table}

\begin{figure}
\begin{center}
\includegraphics[scale=0.35]{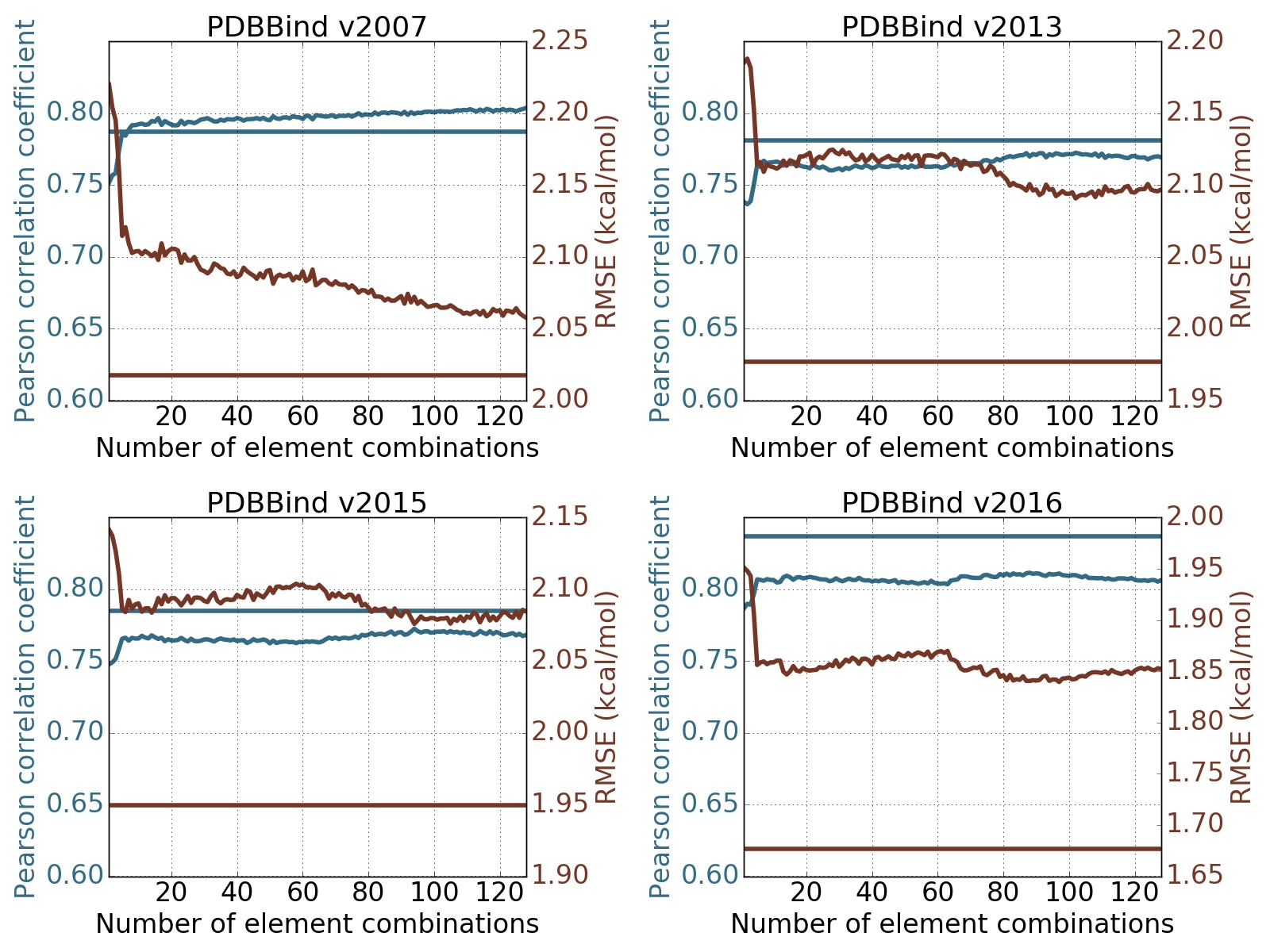}
\end{center}
\caption{The performance of the model against the number of included element combinations. The Betti-1 and Betti-2 barcodes computed with alpha complex is used. Features are generated following \textit{barcode statistics} method. Element combinations are all possible paired choices of one item from  \{ C, N, O, CN, CO, NO, CNO, CNOS \} in protein and another item from \{ C, N, O, S, CN, CO, CS, NO, NS, OS, CNO, CNS, COS, NOS, CNOS, CNOSPFClBrI \} in ligand, which  result in 128 element combinations. The horizontal straight lines represents the performance of the 2D representation with deep convolutional neural network. The blue and red colors correspond to Pearson correlation coefficient and RMSE (kcal/mol) respectively.}\label{fig:FP}
\end{figure}

\paragraph{Robustness of GBT algorithm against redundant  element combination features and potential overfitting}

It is intuitive that combinations of more than 2 element types are able to enrich the representation especially in the case of higher dimensional Betti numbers. However, the consideration of combination of more element types rapidly increases the dimensional of feature space. In the high dimensional feature space, it is almost inevitable that there exists nonessential and redundant features. Additionally, the importance of a feature varies across different problems and data sets. Therefore, it is preferable to keep all the potentially important features in a general model which is expected to cover a wide range of situations. To test the robustness of the model against unimportant features, we select a total of 128 element combinations    (i.e., all possible paired choices of one item from \{C, N, O, CN, CO, NO, CNO, CNOS\} in protein  and another item from \{C, N, O, S, CN, CO, CS, NO, NS, OS, CNO, CNS, COS, NOS, CNOS, CNOSPFClBrI\} in ligand). The Betti-0, Betti-1 and Betti-2 barcodes are computed for all combinations using  alpha complex with Euclidean distance. Features are generated following the barcode statistics method.
 
A general model with all the features is generated in the first place. The element combinations are then sorted according to their importance scores in the general model. Starting from the most important element combination, one element combination is added to the feature vector each time and then the resulting  feature vector is  passed to the machine learning training and testing procedure. The level of adding element combinations is based on their importance scores and thus that a less important feature is added each step.

Figure \ref{fig:FP} depicts the changes of Pearson correlation coefficient and RMSE (kcal/mol) with respect to the increase of element combinations in predicting four PDBBind core sets. In all cases, the inclusion of top combinations can readily deliver very good models. The behavior of the present method in PDBBind v2007 is quite different from that in other data sets. The performance of the present method improves almost monotonically as the element combination increases. However, in other three cases,  the improvement is unsteady. Nevertheless, the performance    fluctuates within a small range, which indicates that the present method is reasonably stable against the increase in element combinations. From a different perspective, the increase in element combinations might lead to overfitting in machine learning. Since the model parameters are fixed before the experiments, it shows that GBT algorithms are not very sensitive to redundant features and are robust against overfitting.

\paragraph{Usefulness of more than 2 element types for interactive Betti-0 barcodes}

While using element combinations with more than 2 element types with higher dimensional Betti numbers enriches characterization of geometry, it remains to assess whether interactive Betti-0 characterization will benefit from element combinations with more element types. As an example, we denote interactive Betti-0 barcodes for carbon and nitrogen atoms from protein and oxygen atoms from ligand by $\mathbf{B}_{\rm{CN-O}}$, barcodes for carbon atoms from protein and oxygen atoms from ligand by $\mathbf{B}_{\rm{C-O}}$, and barcodes for nitrogen atoms from protein and oxygen atoms from ligand by $\mathbf{B}_{\rm{N-O}}$. In the case of persistent homology barcode representation,  $\mathbf{B}_{\rm{CN-O}}$ is not strictly the union of $\mathbf{B}_{\rm{C-O}}$ and $\mathbf{B}_{\rm{N-O}}$. However $\mathbf{B}_{\rm{CN-O}}$ might be redundant to $\mathbf{B}_{\rm{C-O}}$ and $\mathbf{B}_{\rm{N-O}}$. To address this concern, we test features from interactive Betti-0 barcodes with the 36 element combinations (i.e., \{ C, N, O, S \} for protein and \{ C, N, O, S, P, F, Cl, Br, I \} for ligand) and features for the 160 selected element combinations (i.e., \{ C, N, O, S, CN, CO, NO, CNO \} for protein and \{ C, N, O, S, P, F, Cl, Br, I, CN, CO, CS, NO, NS, OS, CNO, CNS, COS, NOS, CNOS \} for ligand), which are listed as feature group 2 and feature group 1 in Table \ref{tab:bfp}. In all the four cases, the features of the 36 combinations (feature group 2) slightly outperforms or performs as well as the features of the 160 combinations (feature group 1) suggesting that element combinations with more than 2 element types are redundant to all the combinations with 2 element types in the case of interactive Betti-0 characterization.
 
\paragraph{Importance of atomic charge information}
 
In element specific persistent homology, atoms of different element types are characterized separately, which offers a rough and implicit description of the electrostatics of the system.  However, such implicit treatment of electrostatics may lose important information because atoms behave differently at different oxidation states. Therefore, we explicitly embed atomic charges in interactive Betti-0 barcodes as described in  Eq. (\ref{eq:chg}). The resulting topological features are given  in feature group 4 in Table \ref{tab:bfp}. It can be seen from Table \ref{tab:bfp} that the combination of feature group 4 and the Euclidean distance based interactive Betti-0 barcodes (listed as feature group 6 and 7) generally outperforms the results obtained with only Euclidean distance based features. This observation suggests that electrostatics play an important role and should be taken care of explicitly for the protein-ligand binding problem. 
 
\paragraph{Relevance of  elements that are rare with respect to the data sets}

\begin{figure}
\begin{center}
\includegraphics[scale=0.25]{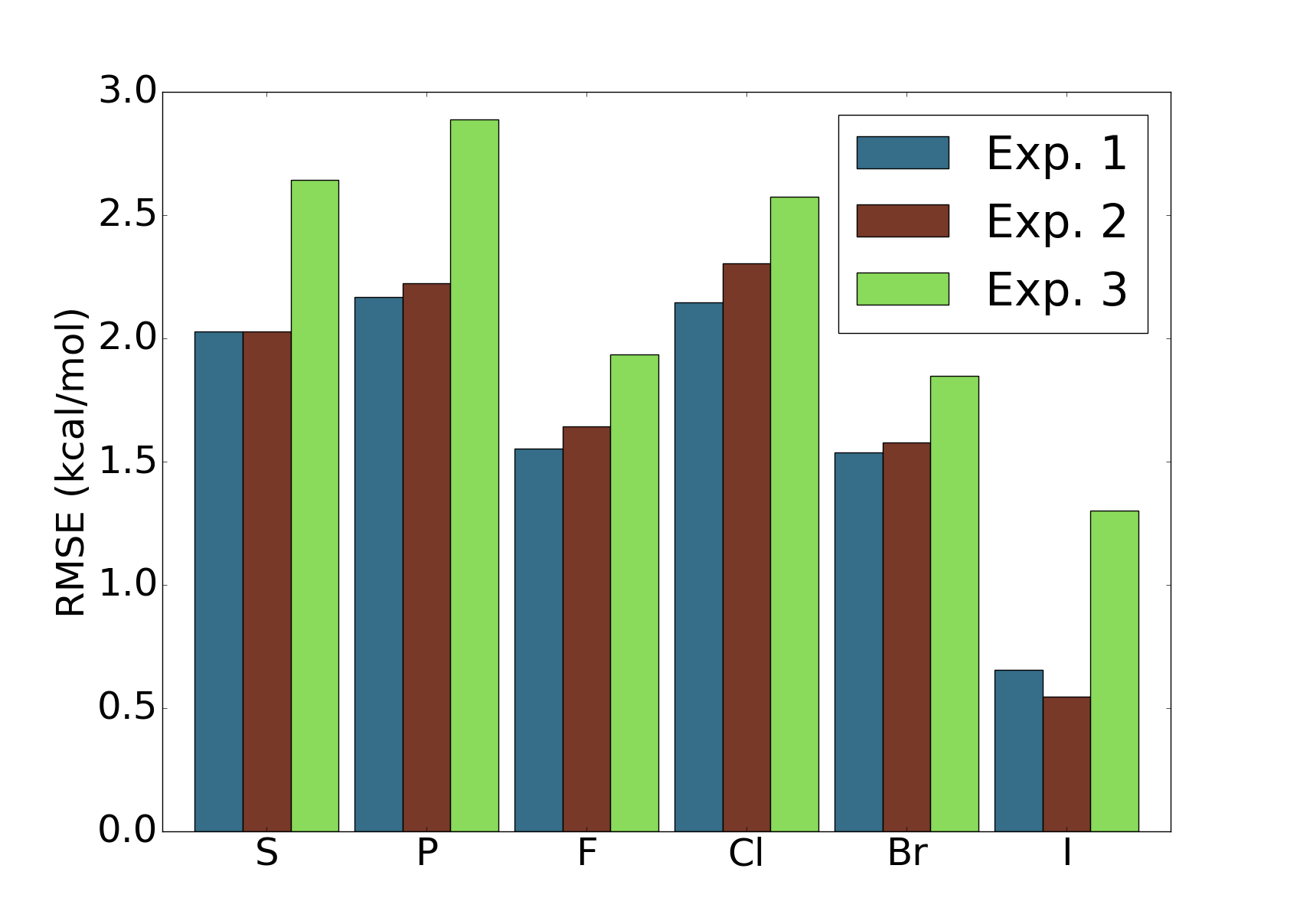}
\caption{Assessment of performance of the model on samples with elements that are rare in the data sets. For the four data sets PDBBind v2007, v2013, v2015, and v2016 \cite{PDBBind:2015}, and for each element, the testing set is the subset of the original core sets with only ligands that contain atoms of the particular element type. The features used are features with ID=7 in Table \ref{tab:bfp}. The reported RMSE is the average taken over the four data sets. Experiment 1: Training set is the original training set and all the features are used. Experiment 2: Training set is the original training set and only features that do not involve the particular element are used. Experiment 3: Training set is the original training set excluding the samples that contain atoms of the particular element type and all features are used. For most of the elements, experiment 1 achieves the best result and experiment 3 yields the worst performance.} 
\label{fig:rareelements}
\end{center}
\end{figure}

Since the majority of the samples in both training and testing sets only contain atoms of element types, C, N, O, and H, the performance of the model on the samples with rare occurring elements with respect to data sets is hardly reflected by the overall performance statistics. For simplicity, we refer to such rarely occurring elements with respect to data sets simply by rarely occurring elements in the discussion follows. To assess the aspects of the model that potentially affect the performance on the samples containing rarely occurring elements, we picked the samples containing each rarely occurring element from the original testing set as a new testing set. Three experiments are carried out to address two questions: ``Are the training samples containing the same rarely occurring element crucial?" and ``Are features addressing the rarely occurring element important?". A short answer is yes to both according to the results shown in Figure \ref{fig:rareelements}. Specifically, for each rarely occurring element, the exclusion of samples containing this element in training set and the exclusion of features addressing this element will both cause degenerated results. It is also shown that the exclusion of samples of the rarely occurring element leads to much worse results. This observation suggests that the same interactions may lead to different binding properties for molecules with different compositions. Since both modifications of the model deliver worse results, we conclude that including the samples in the training set with similar compositions to the test sample is crucial to the success of the model  on this specific test sample.  Even the inclusion of features of more element types or element combinations does not deliver better results in the general testing sets, such features should still be kept in the model in case that a sample with a similar element composition comes in as a test sample.

\paragraph{2D persistence for deep  convolutional neural networks  }

 \begin{figure}
\begin{center}
\includegraphics[scale=0.15]{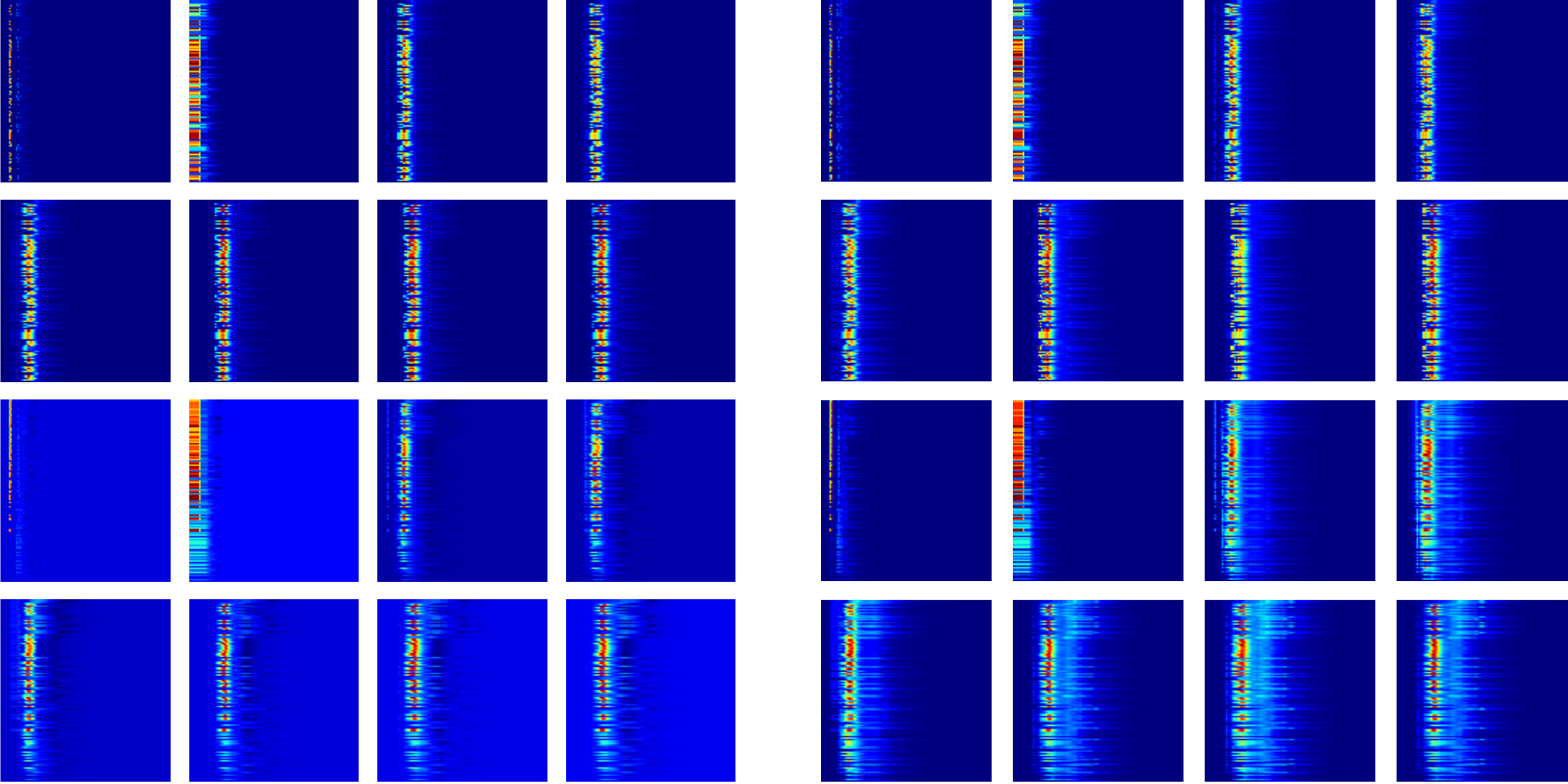}
\end{center}
\caption{The heat map plot of the 16 channels. The mean value (left image) and the standard deviation (right image) of each digit over the PDBBind v2016 refined set are shown. The top 8 maps are for protein-ligand complex and the other 8 maps are for the difference between protein-ligand complex and protein only. For each map, the vertical axis is the     element combinations ordered according to their importance  and the horizontal  axis is the dimension of spatial scales.}\label{fig:feature2d2}
\end{figure}

Deep learning is potentially more powerful than many other machine learning algorithms when the data size is sufficiently large. In the present work, it is natural to construct a 2D  representation  by incorporating the element combination as an additional dimension, resulting  in 16 channels as defined in Section \ref{sec:SDB}. Here 128 element combinations 
(i.e., all possible paired choices of one item from \{C, N, O, CN, CO, NO, CNO, CNOS\} in protein  and another item from \{C, N, O, S, CN, CO, CS, NO, NS, OS, CNO, CNS, COS, NOS, CNOS, CNOSPFClBrI\} in ligand) are used for 2D analysis. The advantage of introducing this extra dimension with convolutional neural networks is to prevent unimportant features from interacting with important ones at the lower levels of the model whilst generally unimportant features are still kept in the model in case that they are essential to specific problems or a certain portion of the data set. Figure \ref{fig:feature2d2} illustrates the mean value and the standard deviation of the PDBBind v2016 refined set. The existence of significant standard deviations for relatively unimportant element combinations indicates that these features might still contribute to the overall prediction.

As shown in Figure \ref{fig:FP}, for all the data sets except the PDBBind v2007 set, the 2D representation with convolutional neural networks performs significantly better. The inferior performance of  convolutional neural networks in v2007 might be   a result of the small data size. Note that v2017 training set has 1102 protein-ligand complexes, whereas other training sets have more than 2700 complexes. Consequently,  deep convolutional neural networks are able to outperform the GBT algorithm in predicting v2013,   v2015 and  v2016 core sets.  Indeed, deep  convolutional neural networks have advantages in dealing with large data sets. 

\subsection{Structure-based virtual screening}\label{Exp:VS}

In this section, we examine the performance of the proposed method for the main application in this paper, which is structure-based virtual screening. The dataset is much larger than the two applications on protein-ligand binding affinity prediction. Therefore, the best performing procedures in ligand-based binding affinity prediction and protein-ligand-complex-based binding affinity prediction are applied in this virtual screening application, since tuning parameters on such big dataset is too time consuming.

\paragraph{DUD data set}
The directory of useful decoys (DUD) \cite{huang2006benchmarking,  mysinger2010rapid} is used to benchmark our topological approach for virtual screening. The DUD data set contains 40 protein targets from six classes, i.e.,  nuclear hormone receptors, kinases, serine proteases, metalloenzymes, folate enzymes, and other enzymes. A total of about 3000 active ligands were identified from literature. The number of ligands for each target ranges from tens to hundreds. At most 36 decoys were constructed for each ligand, from the ZINC database of commercially available compounds \cite{irwin2005zinc}. The decoys were selected so that they possess similar physical properties to the ligands but have dissimilar molecular topology. The physical properties include molecular weight, the log P value, and number of hydrogen bonding groups. This results in a total of about 100000 compounds. A discrepancy between calculated partial charges for the ligand and decoy sets was reported for the original release 2 of DUD datasets, which makes it trivial for virtual screening methods to distinguish between the two sets using those charges. In this work, we use the sets with recalculated Gasteiger charges for both ligand and decoy sets \cite{armstrong2010electroshape}. 

\paragraph{Data processing}
In structure-based virtual screening, the possible complex structures of the target protein and the small molecule candidate are required. For the DUD dataset, the structures of the 40 protein targets, the ligands, and the decoys are given, and we generate protein-ligand complexes or protein-decoy complexes by using docking software. To this end, we first add missing atoms to proteins by using the profix utility in Jackal software package \cite{ZXiang:2001}. The receptors and ligands or decoys are prepared using the scripts prepare\_receptor4.py and prepare\_ligand4.py provided by the AutoDockTools module in MGLTools package (version 1.5.6) \cite{morris2009autodock4}.
The bounding box of the binding site is defined as a cube with edge size equal to 27 \AA, centered at the geometric center of the crystal ligand. AutoDock Vina (version 1.1.2) \cite{Trott:2010AutoDock} is used to dock the ligand or decoy to the receptor. The option exhaustiveness is set to 16 and all the other parameters are set to their default values. In each docking experiment, the pose having the lowest binding free energy reported by AutoDock Vina, is used by the machine learning based model. 

\begin{table}[ht]
\begin{center}
\rowcolors{2}{gray!18}{white}
\begin{tabular}{ll}
\toprule
\rowcolor{gray!50}
Method & Parameters \\
\midrule
GBT & n=2000, s=0.5, cw=100:1, lr=0.01, mf=sqrt \\
RF  & n=2000, cw=balanced\_subsample \\
ET  & n=2000, cw=balanced\_subsample \\
\bottomrule
\end{tabular}
\caption{The parameters used for the ensemble of trees methods while the other parameters are set to default. GBT: gradient boosting trees. RF: random forest. ET: extra trees. n: n\_estimators. s: subsample. cw: class\_weight. lr: learning\_rate. mf: max\_feature}\label{tab:ml}
\end{center}
\end{table}

\begin{table}[ht]
\begin{center}
\rowcolors{2}{gray!18}{white}
\begin{tabular}{lllllll}
\toprule
\rowcolor{gray!50}
 Target & \multicolumn{3}{l}{ADV} & \multicolumn{3}{l}{TopVS-ML} \\
\rowcolor{gray!50}
 & EF$_{2\%}$ & EF$_{20\%}$ & AUC & EF$_{2\%}$ & EF$_{20\%}$ & AUC \\
 \midrule
ACE & 4.1 & 1.4 & 0.42 & 5.1 & 3.2 & \textbf{0.81} \tabularnewline
AChE & 4.7 & 2.8 & \textbf{0.67} & 0.5 & 1.2 & 0.61 \tabularnewline
ADA & 0.0 & 0.4 & 0.49 & 13.0 & 3.7 & \textbf{0.89} \tabularnewline
ALR2 & 2.0 & 2.7 & \textbf{0.74} & 2.0 & 1.4 & 0.67 \tabularnewline
AmpC & 2.4 & 0.2 & 0.34 & 0.0 & 0.2 & \textbf{0.53} \tabularnewline
AR & 17.0 & 3.8 & 0.81 & 19.5 & 4.3 & \textbf{0.90} \tabularnewline
CDK2 & 9.0 & 2.4 & 0.64 & 3.8 & 3.8 & \textbf{0.85} \tabularnewline
COMT & 13.1 & 1.4 & 0.56 & 17.4 & 2.7 & \textbf{0.74} \tabularnewline
COX1 & 9.9 & 2.8 & 0.76 & 9.9 & 3.2 & \textbf{0.84} \tabularnewline
COX2 & 20.7 & 3.9 & 0.86 & 18.7 & 4.9 & \textbf{0.97} \tabularnewline
DHFR & 6.4 & 2.8 & 0.82 & 13.0 & 4.7 & \textbf{0.96} \tabularnewline
EGFr & 3.4 & 1.6 & 0.63 & 19.5 & 4.8 & \textbf{0.96} \tabularnewline
ER$_{{\rm agonist}}$ & 17.8 & 3.3 & \textbf{0.84} & 9.3 & 3.0 & 0.81 \tabularnewline
ER$_{{\rm antagonist}}$ & 10.2 & 2.3 & 0.70 & 0.0 & 3.6 & \textbf{0.86} \tabularnewline
FGFr1 & 0.4 & 0.8 & 0.44 & 10.9 & 4.8 & \textbf{0.95} \tabularnewline
FXa & 1.0 & 1.3 & 0.63 & 1.7 & 4.3 & \textbf{0.89} \tabularnewline
GART & 0.0 & 1.9 & \textbf{0.75} & 2.6 & 0.6 & 0.49 \tabularnewline
GPB & 0.0 & 0.9 & 0.48 & 0.0 & 2.2 & \textbf{0.71} \tabularnewline
GR & 5.7 & 1.2 & 0.57 & 0.0 & 2.6 & \textbf{0.77} \tabularnewline
HIVPR & 5.6 & 2.6 & 0.74 & 4.8 & 4.2 & \textbf{0.90} \tabularnewline
HIVRT & 8.2 & 1.9 & 0.64 & 9.4 & 4.1 & \textbf{0.88} \tabularnewline
HMGR & 0.0 & 0.9 & 0.53 & 23.1 & 5.0 & \textbf{0.98} \tabularnewline
HSP90 & 0.0 & 0.9 & 0.64 & 5.5 & 4.5 & \textbf{0.93} \tabularnewline
InhA & 13.4 & 1.9 & 0.56 & 22.7 & 4.5 & \textbf{0.95} \tabularnewline
MR & 16.7 & 4.0 & 0.82 & 6.7 & 4.7 & \textbf{0.89} \tabularnewline
NA & 0.0 & 0.3 & 0.37 & 1.0 & 3.5 & \textbf{0.84} \tabularnewline
P38 MAP & 1.4 & 1.7 & 0.59 & 17.7 & 4.5 & \textbf{0.95} \tabularnewline
PARP & 4.2 & 2.7 & 0.71 & 0.0 & 2.4 & \textbf{0.74} \tabularnewline
PDE5 & 8.0 & 1.9 & 0.61 & 6.9 & 3.5 & \textbf{0.86} \tabularnewline
PDGFrb & 3.5 & 0.5 & 0.32 & 20.3 & 4.9 & \textbf{0.96} \tabularnewline
PNP & 0.0 & 0.7 & 0.59 & 8.9 & 4.1 & \textbf{0.90} \tabularnewline
PPARg & 17.7 & 3.4 & \textbf{0.82} & 0.6 & 1.6 & 0.72 \tabularnewline
PR & 1.9 & 1.1 & 0.52 & 9.4 & 4.0 & \textbf{0.88} \tabularnewline
RXRa & 28.2 & 4.8 & \textbf{0.95} & 7.7 & 2.5 & 0.79 \tabularnewline
SAHH & 10.4 & 3.0 & 0.80 & 1.5 & 4.2 & \textbf{0.84} \tabularnewline
SRC & 5.6 & 2.3 & 0.71 & 20.2 & 4.9 & \textbf{0.97} \tabularnewline
thrombin & 8.3 & 2.6 & 0.72 & 2.8 & 2.2 & \textbf{0.75} \tabularnewline
TK & 0.0 & 0.9 & 0.56 & 6.9 & 2.5 & \textbf{0.64} \tabularnewline
trypsin & 3.1 & 1.9 & 0.58 & 0.0 & 2.0 & \textbf{0.74} \tabularnewline
VEGFr2 & 10.2 & 2.2 & 0.63 & 21.0 & 4.7 & \textbf{0.96} \tabularnewline
\bottomrule
Average & 6.9 & 2.0 & 0.64 & 8.6 & 3.4 & \textbf{0.83} \tabularnewline
\bottomrule
\end{tabular}
\caption{The median results of 10 repeated runs with different random seeds are reported. The best AUC in each row is marked in bold. The second block of AutoDock Vina (ADV) results are acquired from our ADV runs. The number of ligands for each target is listed in the parenthesis next the target name.}\label{tab:DUDdetail}
\end{center}
\end{table}

\paragraph{Evaluation }

Two metrics, the enrichment factor (EF) and the area under the receiver operating characteristic curve (AUC), are used to evaluate each method's ability of discriminating ligands from decoys. The AUC is defined as 
\begin{equation}\label{eq:auc}
{\rm AUC} = 1-\frac{1}{N_a}\sum\limits_{i=1}^{N_a}\frac{N^i_d}{N_d},
\end{equation}
where $N_a$ is the number of active ligands, $N_d$ is the total number of decoys, and $N^i_d$ is the number of decoys that are higher ranked than the $i$th ligand. An AUC value of 0.5 is the expected value of random selection, whereas a perfect prediction results in an AUC of 1. The EF at $x$\% denoted by EF$_{x\%}$ evaluates the quality of the set of top $x\%$ ranked compounds, by comparing the percentage of actives in the top $x\%$ ranked compounds to the percentage of actives in the entire compound set. It is defined as 
\begin{equation}\label{eq:EF}
{\rm EF}_{x\%} = \frac{N_a^{x\%}}{N^{x\%}}\cdot\frac{N}{N_a},
\end{equation}
where $N_a^{x\%}$ is the number of active ligands in the top $x\%$ ranked compounds, $N^{x\%}$ is the number of top $x\%$ ranked compounds, $N$ is the total number of compounds, and $N_a$ is the total number of active ligands.

\paragraph{Topology based machine learning models}

Unlike physical based models, a training set of sufficient size is needed for machine learning base models. To evaluate the performance of various methods on the DUD data set, the structure data obtained from docking associated with one protein target are used as the test set each time \cite{pereira2016boosting}. For the selection of the training set of a given protein target, we follow a procedure given in the   literature \cite{huang2006benchmarking}, where the entries associated to the rest of the proteins, excluding those that are within the same class of the testing protein and those that have reported positive cross-enrichment with the testing protein, are taken as the training set. The 40 proteins are split into 6 classes \cite{arciniega2014improvement}. A detailed list of proteins that are excluded from the training set of each protein is given in Supplementary Table S6.

\begin{table}[ht]
\begin{center}
\rowcolors{2}{gray!18}{white}
\begin{tabular}{lll}
\toprule
\rowcolor{gray!50}
Method & AUC & Ref\\
\midrule
TopVS-ML             & 0.83 & \\
DeepVS-ADV           & 0.81 & \cite{pereira2016boosting} \\
ICM$^a$              & 0.79 & \cite{neves2012docking} \\
NNScore1-ADV$^b$     & 0.78 & \cite{durrant2013comparing} \\
Glide SP$^a$         & 0.77 & \cite{cross2009comparison} \\
DDFA-ALL             & 0.77 & \cite{arciniega2014improvement} \\
DDFA-RL              & 0.76 & \cite{arciniega2014improvement} \\
NNScore2-ADV$^b$     & 0.76 & \cite{durrant2013comparing} \\
DDFA-ADV             & 0.75 & \cite{arciniega2014improvement} \\
DeepVS-Dock          & 0.74 & \cite{pereira2016boosting} \\
DDFA-AD4             & 0.74 & \cite{arciniega2014improvement} \\
Glide HTVS$^b$       & 0.73 & \cite{durrant2013comparing} \\
Surflex$^a$          & 0.72 & \cite{cross2009comparison} \\
Glide HTVS           & 0.72 & \cite{cross2009comparison} \\
ICM                  & 0.71 & \cite{neves2012docking} \\
RAW-ALL              & 0.70 & \cite{arciniega2014improvement} \\
AutoDock Vina$^b$    & 0.70 & \cite{durrant2013comparing} \\
Surflex              & 0.66 & \cite{cross2009comparison} \\
Rosetta Ligand       & 0.65 & \cite{arciniega2014improvement} \\
AutoDock Vina        & 0.64 & \cite{arciniega2014improvement} \\
ICM                  & 0.63 & \cite{cross2009comparison} \\
FlexX                & 0.61 & \cite{cross2009comparison} \\
Autodock4.2          & 0.60 & \cite{arciniega2014improvement} \\
PhDOCK               & 0.59 & \cite{cross2009comparison} \\
Dock4.0              & 0.55 & \cite{cross2009comparison} \\
\bottomrule
\end{tabular}
\caption{$^a$Tuned by expert knowledge. $^b$Determined using a different data set of decoys.}\label{tab:DUDcompare}
\end{center}
\end{table}

Our topology based machine learning  model, called \textit{TopVS-ML}, relies on manually constructed features and utilizes ensemble of trees methods. For the complex with the small molecules (i.e., ligands and decoys) docked to the receptor, features  {R-B0-I-BP}, {R-B0-CI-S}, and {A-B12-E-S} are used, whereas features {R-B012-M1-S} and  {A-B012-E-S} are used for the small molecules. The gradient boosting trees method, random forest method, and extra trees method are employed as voters. The averaged probabilities output by the three methods are used for the classifier to decide the class of the testing samples. The modules \textit{GradientBoostingClassifier}, \textit{RandomForestClassifier}, and \textit{ExtraTreesClassifier} in the scikit-learn package \cite{scikit-learn} are used.  The parameters for the three modules are listed in Table \ref{tab:ml}. The performance on each of 40 protein targets is reported in Table \ref{tab:DUDdetail}. We  have  also generated virtual screening results of AutoDock Vina (ADV) based on the modeled binding free energy and compared them with those of the present TopVS-ML in terms of enrichment factors and the areas under the receiver operating characteristic curve (AUC). 
  A comprehensive comparison of average AUC with those from a large number of methods is given in Table \ref{tab:DUDcompare}.

In our final model reported in Table \ref{tab:DUDcompare}, we use topological descriptors of both protein-compound interactions and only the compounds (i.e., ligands and decoys). We have also tested our models using either one of the aforementioned descriptions. When only topological descriptor of small molecules are used, which falls into the category of ligand-based virtual screening,  an AUC of 0.81 is achieved. For the topological model using only the descriptions of protein-ligand interactions,  an AUC of 0.77 are achieved. 
An AUC of 0.83 is obtained with a model combining both sets of descriptors which is better than each individual performance, suggesting that the two groups of descriptors are complementary to each other and are both important for achieving satisfactory results. The marginal improvement made by protein-compound complexes maybe due to the various docking quality. Similar situation was encountered by a deep learning method \cite{pereira2016boosting}. For the targets with high quality results by Autodock Vina (AUC of ADV > 0.8), the ligand-based features achieve an AUC of 0.81 and the complex-based features achieve an AUC of 0.86. On the other hand, for the targets with low quality results by Autodock Vina (AUC of ADV < 0.5), the ligand-based features achieve an AUC of 0.81 and the complex-based features achieve an AUC of 0.74. The results of these cases are listed in Table S8 and S9. This observation suggests that the performance of features describing the interactions and the geometry of protein-compounds complexes highly depends on the quality of docking results.

Our  model with small molecular  descriptors  delivers an AUC of 0.81, which is comparably well to the other top performing methods. The performance of this model is also competitive in the regime of protein-ligand binding affinity prediction based on experimentally solved complex structures as is shown in Section \ref{Exp:LigandBased}. These results suggest that topology based small molecule characterization proposed in this work is potentially useful in other applications to small molecules, such as predictions of toxicity, solubility and partition coefficient. 

\section{Conclusion}
Persistent homology is a relatively new branch of algebraic topology and is the main workhorse in topological data analysis. The topological simplification of biomolecular systems was a major motivation of the earlier persistent homology development  \cite{Edelsbrunner:2002,Zomorodian:2005}. Persistent homology has been applied to computational biology \cite{Kasson:2007,Gameiro:2014,Dabaghian:2012,Perea:2015b,Gameiro:2014}, including our efforts \cite{KLXia:2014c,KLXia:2015a,KLXia:2015d,KLXia:2015e,KLXia:2015b,BaoWang:2016a,ESES:2017}. However, the predictive power of primitive persistent homology was limited in  early applications \cite{ZXCang:2015}.  To address this challenge, we have recently introduced element specific persistent homology to retain  chemical and biological information during the topological abstraction of biomolecules \cite{ZXCang:2017a,ZXCang:2017b,ZXCang:2017c}. This approach offers competitive predictions of protein-ligand binding affinity and mutation induced protein stability changes. However, its representability and predictive power for small molecules and their interaction with macromolecules remain unknown. 

The present work further introduces multicomponent persistent homology, multi-level persistent homology and electrostatic persistence for  chemical and biological characterization, analysis and modeling. Multicomponent persistent homology takes a  combinatorial approach to create possible element specific topological representations.  Multi-level persistent homology allows tailored topological descriptions of any desirable interaction in biomolecules.  Electrostatic persistence incorporates partial charges that are essential to biomolecules in topological invariants.  These approaches are implemented via the appropriate construction  of the distance matrix for filtration.  The representation power and  reduction power of multicomponent persistent homology, multi-level persistent homology and electrostatic persistence are validated by two databases, namely PDBBind \cite{PDBBind:2015} and DUD \cite{huang2006benchmarking,  mysinger2010rapid}.  PDBBind involves more than 4,000 protein-ligand complexes and and DUD contains near 100,000 small compounds.   Two classes of problems are used to test the proposed topological methods, including the regression (prediction) of protein-ligand binding affinities and the discrimination of active ligands from non-active decoys (virtual screening). In both problems, we examine the representability of proposed topological methods on small molecules, which are somewhat more difficult to describe by persistent homology due to their chemical diversity, variability and sensitivity. Additionally,  these methods are tested on their ability to handle the full protein-ligand complexes.   Advanced  machine learning methods, including  Wasserstein metric based k nearest neighbors (KNNs), gradient boosting trees (GBTs),  random forest  (RF), extra trees (ETs) and deep convolutional neural networks (CNNs) are utilized in the present work to facilitate the proposed topological methods in quantitative  biomolecular predictions. The thorough examination of the method on the prediction of binding affinity for experimentally solved protein-ligand complexes leads to a structure-based virtual screening method, TopVS, which outperforms other modern methods. The feature sets introduced in this work for small molecules and protein-ligand complexes can be extended to other applications such as 3D-structure based prediction of toxicity, solubility, and partition coefficient for small molecules and complex structure based prediction of protein-nucleic acid binding and protein-protein binding affinities.

\vspace{1cm}
%
%
%
%
%
%
%
%
%
%

\section*{Acknowledgment} 
This work was supported in part by NSF Grants DMS-1721024  and IIS-1302285     and
MSU Center for Mathematical Molecular Biosciences Initiative.

\vspace{1cm}
\bibliographystyle{ieeetr}

\end{document}